\documentclass[
  journal=pasa,
  manuscript=research-paper, 
  year=2020,
  volume=37,
]{cup-journal}

\usepackage{microtype,siunitx,booktabs}
\usepackage{graphicx}
\usepackage{pdflscape}
\usepackage{amssymb}
\usepackage{amsmath}
\usepackage{hyperref}
\usepackage{ulem}
\newcommand\micron{\rm{\upmu m}}
\colorlet{juan}{Fuchsia!50!WildStrawberry}
\def\micron{\hbox{$\mu$m}}
\newcommand\ion[2]{\text{#1\,\textsc{\lowercase{#2}}}}	

\title{Galaxy evolution through infrared and submillimeter spectroscopy:\\ Measuring star formation and black hole accretion with JWST and ALMA}

\author{Sabrina Mordini}
\affiliation{Dipartimento di Fisica, Universit\`a di Roma La Sapienza, P.le A. Moro 2, I--00185 Roma, Italy}
\alsoaffiliation{Istituto di Astrofisica e Planetologia Spaziali (INAF--IAPS), Via Fosso del Cavaliere 100, I--00133 Roma, Italy}
\email[S. Mordini]{sabrina.mordini@inaf.it}

\author{Luigi Spinoglio}
\affiliation{Istituto di Astrofisica e Planetologia Spaziali (INAF--IAPS), Via Fosso del Cavaliere 100, I--00133 Roma, Italy}

\author{Juan Antonio Fern\'andez-Ontiveros}
\alsoaffiliation{Istituto di Astrofisica e Planetologia Spaziali (INAF--IAPS), Via Fosso del Cavaliere 100, I--00133 Roma, Italy}
\affiliation{Centro de Estudios de F\'isica del Cosmos de Arag\'on (CEFCA), Plaza San Juan 1, E--44001, Teruel, Spain}

\doi{10.1017/pasa.2020.32}

\received {dd Mmm YYYY}
\revised  {dd Mmm YYYY}
\accepted {dd Mmm YYYY}
\published{dd Mmm YYYY}

\keywords{galaxies: active – galaxies: evolution – galaxies: star formation – infrared: galaxies – techniques: spectroscopic – telescopes} 

\begin{document}

\begin{abstract}
Rest-frame mid- to far-infrared (IR) spectroscopy is a powerful tool to study how galaxies formed and evolved, because a major part of their evolution occurs in heavily dust enshrouded environments, especially at the so-called Cosmic Noon ($1< z < 3$). 
Using the calibrations of IR lines and features, recently updated with {\it Herschel} and {\it Spitzer} spectroscopy, we predict their expected fluxes with the aim to measure the Star Formation (SF) and the Black Hole Accretion (BHA) rates in intermediate to high redshift galaxies.

On the one hand, the recent launch of the \textit{James Webb Space Telescope} (\textit{JWST}) offers new mid-IR spectroscopic capabilities that will enable for the first time a detailed investigation of both the SF and the BHA obscured processes as a function of cosmic time. We make an assessment of the spectral lines and features that can be detected by JWST-MIRI in galaxies and active galactic nuclei up to redshift $z \sim 3$. The fine-structure lines of [MgIV]4.49$\mu$m and [ArVI]4.53$\mu$m can be used as BHA rate tracers for the $1 \lesssim z \lesssim 3$ range, and we propose the [NeVI]7.65$\mu$m line as the best tracer for $z \lesssim 1.5$. The [ArII]6.98$\mu$m and [ArIII]8.99$\mu$m lines can be used to measure the SF rate at $z \lesssim 3$ and $z \lesssim 2$, respectively, while the stronger [NeII]12.8$\mu$m line exits the JWST-MIRI spectral range above $z \gtrsim 1.2$. At higher redshifts, the PAH features at 6.2$\mu$m and 7.7$\mu$m can be observed at $z \lesssim 3$ and $z \lesssim 2.7$, respectively.

On the other hand, rest-frame far-IR spectroscopic observations of high redshift galaxies ($z \gtrsim 3$) have been collected with the Atacama Large Millimeter Array (ALMA) in the last few years. The observability of far-IR lines from ALMA depends on the observed frequency, due to the significant decrease of the atmospheric transmission at the highest frequencies ($\gtrsim 420\, \rm{Hz}$). The [CII]158$\mu$m line is a reliable tracer of the SF rate and can in most cases ($0.9 \lesssim z \lesssim 2$ and $2 \lesssim z \lesssim 9$) be observed. Additionally, we propose the use of the combination of [OIII]88$\mu$m and [OI]145$\mu$m lines as an alternative SF rate tracer, that can be detected above $z \gtrsim 3$.

Overall, we emphasize the importance of using multi-feature analysis to measure both BHA and SFR, since individual tracers can be strongly dependent on the local ISM conditions and vary from source to source.

However, we conclude that the peak of the obscured SF and BHA activities at Cosmic Noon falls outside the wavelength coverage of facilities currently operating or under development. A new IR space telescope covering the full IR spectral range from about 10$\mu$m to 300$\mu$m and actively cooled to achieve high sensitivity, will be needed.
\end{abstract}

\section{Introduction}
\label{sec:intro}
The formation and evolution of galaxies is one of the main fields of study in modern astrophysics, nevertheless it has serious limitations due to our lack of knowledge concerning the main processes that drive the evolution. These are the two sources of energy in galaxies, i.e. star formation with the subsequent stellar evolution, and the accretion onto supermassive black holes in active galactic nuclei (AGN). The bulk of activity, however, takes place in heavily dust obscured environments during the so-called Cosmic Noon ($z \sim 1-3$), where almost 90$\%$ of optical and UV radiation is absorbed by dust and re-emitted at longer wavelengths \citep{madau2014,heckman2014}, then followed by a steep decline toward the present epoch. Thus, to fully characterize the various phenomena driving galaxy evolution and their interplay is paramount to explore galaxies throughout their dust obscured phase \citep[e.g.][]{spinoglio2017,spinoglio2021}.

The mid- to far-infrared (IR) spectral range is populated by a wealth of lines and features which offer an ideal tool to probe the dust hidden regions where the bulk of activity in galaxies takes place \citep[e.g.][]{spinoglio1992,armus2007}. The various fine-structure lines in this range cover a wide range of physical parameters such as density, ionization and excitation, and can be used to discriminate between the different conditions of the interstellar medium and the processes taking place there. Most of these lines, however, can be observed only from space IR telescopes, while only a limited number of transitions are detectable by ground-based facilities. In particular, AGN activity can be traced by high-ionization lines, while low to intermediate ionization fine-structure lines are good tracers of star formation processes.

\begin{figure*}
    \centering
    \includegraphics[width=0.5\textwidth]{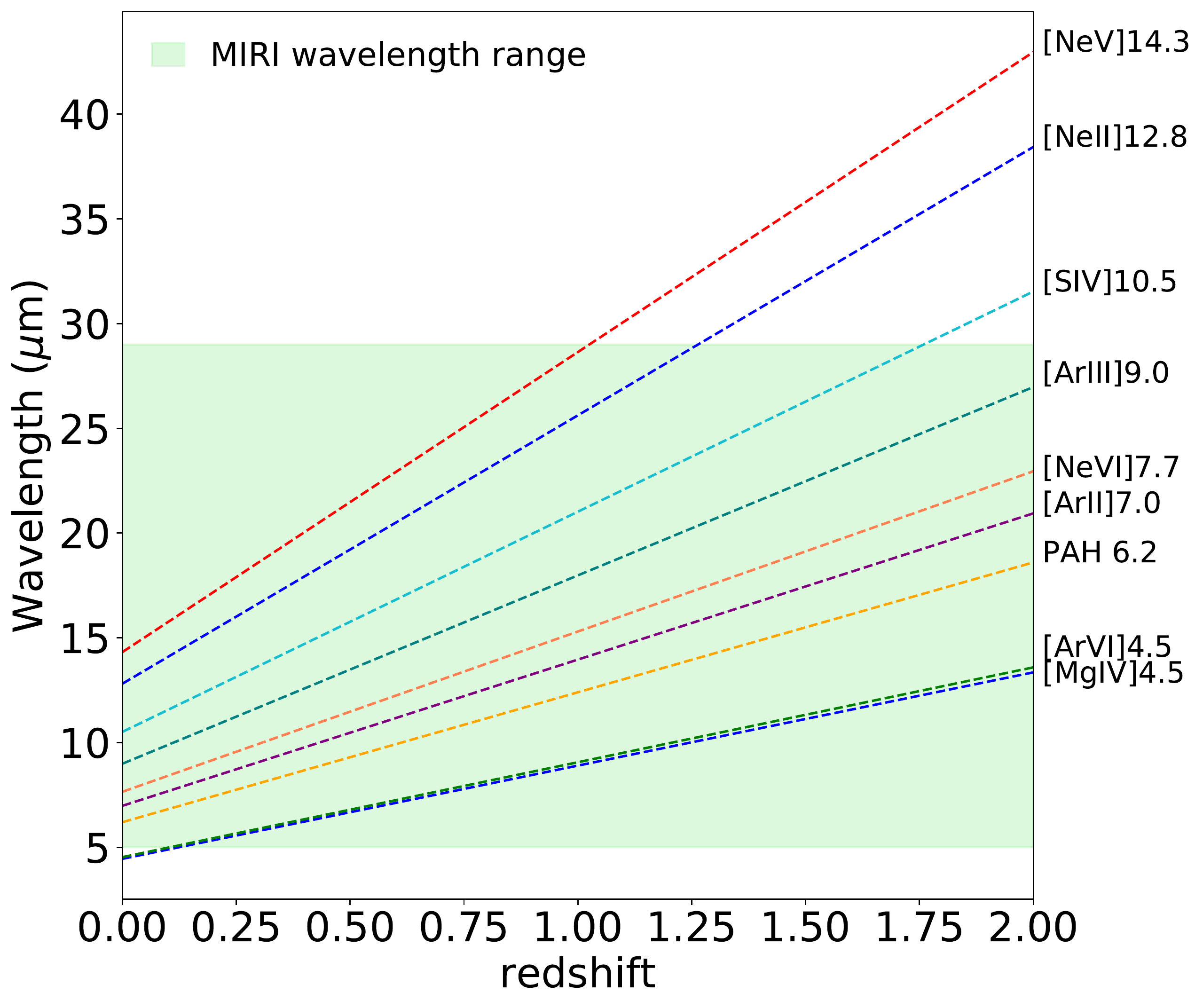}~
    \includegraphics[width=0.5\textwidth]{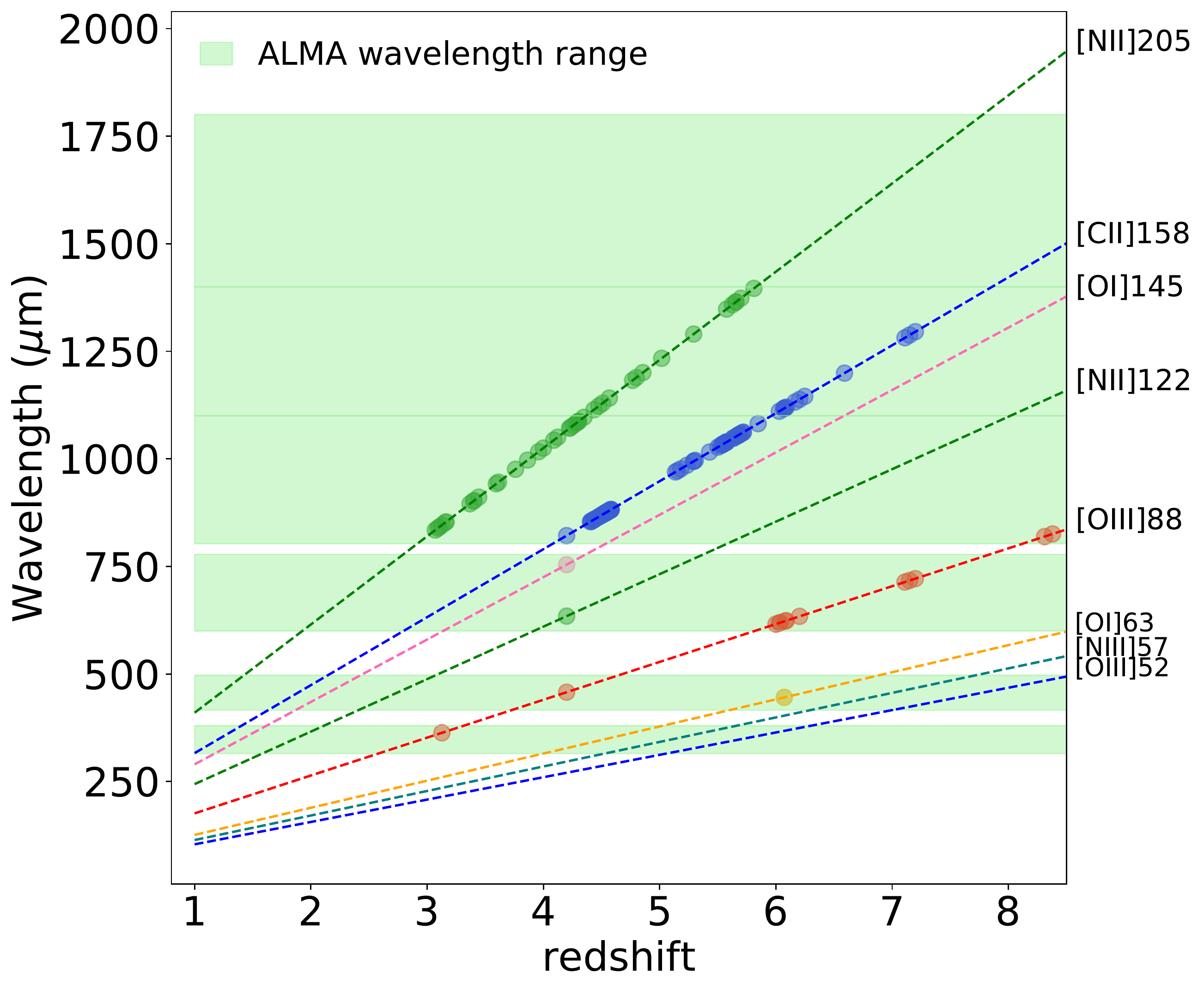}
    \caption{\textit{Left:} observability of key mid-IR (dashed lines) lines as a function of redshift. The shaded area shows the redshift interval covered by the \textit{JWST}-MIRI instrument spectral range. \textit{Right:} observability of key far-IR (dashed lines) lines compared to the ALMA bands (shaded area) in the 1-8 redshift interval. Dots represent current detection for each line. Figure adapted from \citet{mordini2021}.}
    \label{fig:miri_alma_spectral_range}
\end{figure*}

The \textit{James Webb Space Telescope} \citep[\textit{JWST},][]{gardner2006}, successfully launched in December 2021, will observe the Universe in the near- to mid-IR range. In particular, the Mid-InfraRed Instrument \citep[MIRI, ][]{rieke2015,wright2015} will be able to obtain imaging and spectra with unprecedented sensitivity in the 4.9–28.9$\mu$m range. In the near future, the extremely large ground-based telescopes, currently under construction, will have dedicated instruments to obtain spectroscopic observations in the \textit{N} band atmospheric window at 8–13$\mu$m, and thus will be able to measure the [SIV]10.5$\mu$m and [NeII]12.8$\mu$m fine-structure lines, and the 11.3$\mu$m PAH band in local galaxies. irst light from the ESO Extremely Large telescope \citep[E-ELT;][]{gilmozzi2007} is expected by the end of 2025, with the Mid-infrared ELT Imager and Spectrograph \citep[METIS;][]{brandl2021} covering the \textit{N} band with both imaging and low resolution spectroscopy; the Thirty Meter Telescope \citep[TMT;][]{schock2009} {is expected to} be ready for science by the end of 2027, and the Giant Magellan Telescope \citep[GMT;][]{johns2012} will be operational in 2029.

At the present time, only the \textit{Stratospheric Observatory For Infrared Astronomy} \citep[\textit{SOFIA};][]{gehrz2009}, observing from the high atmosphere, can cover the full mid- to far-IR spectral range and detect {line emission from} the brightest galaxies in the local Universe, although with limitations {due to} atmospheric absorption and emission. An example of successful \textit{SOFIA} spectroscopic observations of the [OIII]52$\mu$m and [NIII]57$\mu$m far-IR lines in local galaxies, aimed at the {measurement of chemical abundances}, can be found in \citet{spinoglio2021b}. At much longer wavelengths, in the submillimeter domain, the Atacama Large Millimeter Array \citep[ALMA;][]{wootten2009} can observe within the submillimeter atmospheric bands centered at $\sim$ {0.35\,mm and 0.45\,mm, and the 0.6-3.6\,mm range}. However, the shorter wavelength bands are accessible only when the precipitable water vapour content of the atmosphere is low {enough} (PWV $\lesssim$0.5\,mm) to obtain a peak transmission {of about} 50\%.

In this paper we assess the detection of star formation and BHA tracers, using the JWST-MIRI spectrometer and the ALMA telescope, by computing the predicted fluxes of key lines as a function of redshift, comparing them with the sensitivity of the two instruments. The work is organized as follows. Sections \ref{sec:jwst} and \ref{sec:alma} briefly describe the capabilities of the JWST-MIRI instrument and the ALMA observatory, respectively, for studying galaxy evolution. Section \ref{sec:methods} describes the approach used to derive the predicted line fluxes. Section \ref{sec:results} reports our results, in particular the predictions {for JWST are analysed} in Section \ref{sec:results_jwst}, {while those for} ALMA are in Section \ref{sec:results_alma}. In Section \ref{sec:sfr_bhar} we {provide} the prescriptions to measure the SFR and the BHAR with JWST and ALMA. {These results are discussed} in Section \ref{sec:discussion}, and the main conclusions {of this work} are presented in Section \ref{sec:conclusions}.

\subsection{The Cosmic Noon as seen by JWST in the mid-IR}
\label{sec:jwst}

\begin{table}[t]
\caption{The fine-structure lines in the mid- to far-IR range used in this work. For each line, the columns give: central wavelength, frequency, ionization potential, excitation temperature and critical density. Critical densities and excitation temperatures are from: \citet{launay1977,tielens1985,greenhouse1993,sturm2002,cormier2012,goldsmith2012,farrah2013,satyapal2021}. Adapted from \citet{mordini2021}.}\label{tab:line_properties}
\begin{threeparttable}
\begin{tabular}{lccccc}
\bf Line & \bf $\lambda$ & \bf $\nu$ & \bf I.P. & \bf E & \bf $n_{\rm cr}$ \\
 & ($\, \mu$m) & (GHz) & (eV) & (K) & ($\rm{cm^{-3}}$) \\ \hline
$[\rm{MgIV}]$ & 4.49 & 66811.78& 80 &--& 1.258$\times$10$^{7}$ \\
$[\rm{ArVI}]$ & 4.53 & 66208.58& 75 &-- & 7.621$\times$10$^{5}$ \\
$[\rm{ArII}]$ & 6.98 &42929.48& 16 &-- & 4.192$\times$10$^{5}$ \\
$[\rm{NeVI}]$ & 7.65 & 39188.56 & 126.21 & 1888 & 2.5$\times$10$^5$ \\
$[\rm{ArIII}]$ & 8.99 & 33351.11& 28 & -- & 3.491$\times$10$^{5}$ \\
$[\rm{SIV}]$ & 10.51 & 28524.50 & 34.79 & 1369 & 5.39$\times$10$^4$ \\
$[\rm{NeII}]$ & 12.81 & 23403.00 & 21.56 & 1123 & 7.00$\times$10$^5$ \\
$[\rm{NeV}]$ & 14.32 & 20935.23 & 97.12 & 1892 & 3$\times$10$^4$ \\
$[\rm{OI}]$ & 63.18 & 4744.77 & -- & 228 & 4.7$\times$10$^5$\tnote{a} \\
$[\rm{OIII}]$ & 88.36 & 3393.01 & 35.12 & 163 & 510 \\
$[\rm{NII}]$ & 121.90 & 2459.38 & 14.53 & 118 & 310 \\
$[\rm{OI}]$ & 145.52 & 2060.07 & -- & 98 & 9.5$\times$10$^4$\tnote{a} \\
$[\rm{CII}]$ & 157.74 & 1900.54 & 11.26 & 91 & 20, [2.2\tnote{a} , 4.4\tnote{b} \ ]$\times$10$^3$ \\
$[\rm{NII}]$ & 205.3 & 1460.27 & 14.53 & 70 & 48\\
 \hline
\end{tabular}%
\begin{tablenotes}
    \item[a] Critical density for collisions with hydrogen atoms.
    \item[b] Critical density for collisions with H$_2$ molecules.
\end{tablenotes}
\end{threeparttable}
\end{table}

The JWST-MIRI instrument can perform spectroscopic imaging observations in the 4.9–28.9 $\mu$m range through {four} integral field units. The left panel in Fig. \ref{fig:miri_alma_spectral_range} shows the instrument spectral range and the maximum redshift at which {the main} key lines and features in the mid-IR range can be observed. The JWST-MIRI cut-off wavelength of $\sim$29$\mu$m significantly limits the possibility of probing accretion phenomena {using} the brightest {high-excitation} lines, {which are beyond reach at} $z > 1$, missing about half of the {cosmic} accretion history. The best BHAR tracers in the mid-IR are [\ion{Ne}{v}]$14.3,24.3\, \rm{\micron}$ and [\ion{O}{iv}]$25.9\, \rm{\micron}$ \citep[see, e.g.][]{sturm2002,melendez2008,tommasin2008,tommasin2010}. The [NeV]24.3 $\mu$m and [OIV]25.9 $\mu$m lines are not shown in the left panel of Fig. \ref{fig:miri_alma_spectral_range} due to the limited range available: the former can be observed up to redshift $z \simeq 0.15$, whereas the latter {escapes above} $z \simeq 0.1$. The [NeV]14.3 $\mu$m line can be observed by JWST-MIRI up to redshift $z \simeq 1$.

Alternative BHAR tracers that could be {exploited by} JWST-MIRI beyond $z \sim 1$ have been {recently} proposed by \citet{satyapal2021}. These authors {use photo-ionization simulations to investigate possible spectral diagnostics that separate the AGN and star formation contribution in galaxies using} MIRI and NIRSpec \citep[Near Infrared Spectrograph;][]{bagnasco2007,birkmann2016}. In particular, they {propose} [MgIV]4.49 $\mu$m, [ArVI]4.53$\mu$m and [NeVI]7.65$\mu$m {as the best} tracers {to identify} heavily obscured AGN with moderate {luminosities} up to $z \sim 3$. These lines, however, have not been extensively observed so far, and {thus the predictions are completely based on} photo-ionization models.

{On the other hand}, star {formation can be traced} up to $z \sim 2$ using PAH features, {nevertheless the} behaviour {of these features in high redshift galaxies} is not yet well understood. These {aromatic compounds} could in fact be destroyed by strong radiation fields and/or be less efficiently formed {in} low metallicity {environments}, \citep[e.g. ][]{engelbracht2008,cormier2015,galliano2021}. While alternative SFR tracers {based} in optical lines \citep[e.g.][]{alvarezmarquez2019} or {continuum band fluxes} \citep[e.g.][]{senarath2018} {could be used, the potential of} mid-IR spectroscopic tracers {is a far more relevant for the study of dust-obscured galaxies \citep{ho2007}}.

\subsection{Galaxy evolution with ALMA beyond the Cosmic Noon}
\label{sec:alma}

While JWST will allow us to probe galaxies at redshift below $z \lesssim 3$ using mid-IR lines, the ALMA telescope {has already studied a number of galaxies} at $z \gtrsim 3$. In particular, the SFR {can} be traced using far-IR lines, i.e. [OIII]88$\mu$m and [CII]158$\mu$m up to $z \simeq 8$, as shown in the right panel of Fig. \ref{fig:miri_alma_spectral_range}. An analogous figure can be found in \citet[][see their fig.\,1]{carilli2013}, where the CO transitions and other key tracers of the ISM are shown as a function of redshift versus frequency in terms of the observability by ALMA and JVLA \citep[Karl J. Jansky Very Large Array,][]{perley2011}. As can be seen in the right panel of Fig. \ref{fig:miri_alma_spectral_range}, the [OIII]88$\mu$m, [CII]158$\mu$m and [NII]205$\mu$m lines have been detected over an extended redshift interval, from $z \sim 3-4$ to $z \sim 6-8$, while the shorter wavelength lines such as [OI]63$\mu$m show very few {observations}. No detections have been reported so far for the [OIII]52$\mu$m and [NIII]57$\mu$m lines {with ALMA}. This is due to the {technical} difficulty of observing at {the highest frequencies, where} the atmospheric absorption {limits the available observing time to a} $\sim$10$\%$ of the total. While ALMA cannot detect these far-IR lines in galaxies at the peak of their activity ($1 \lesssim z \lesssim 3$), it allows us to probe galaxies at earlier epochs, thus shedding light on the processes that led to the Cosmic Noon. {Observations of the rest-frame far-IR continuum in galaxies after the reionization epoch} suggest that {large amounts of} dust were already present in these objects \citep[e.g.][]{laporte2017}, and {therefore} IR tracers {will be required} in order to probe {the ISM} and understand the physical processes taking place {in those galaxies}.

\begin{table}[t]
\caption{Line ratios derived from CLOUDY simulations by \citet{fernandez2016} for AGN, SFG and LMG: log(U) indicates the logarithm of the ionization parameter, while log(n$_{H}$) indicates the logarithm of the hydrogen density.}
\label{tab:line_ratios}
\resizebox{\textwidth}{!}{%
\begin{tabular}{lc|cccc}
\hline
\bf AGN & \multicolumn{1}{r|}{log\,U} & -1.5 & -1.5 & -2.5 & -2.5 \\
 & \multicolumn{1}{r|}{log(n$_{H}$/cm$^{-3}$)} & 2 & 4 & 2 & 4 \\[0.1cm] \cline{3-6} 
 & {[MgIV]}4.49/[NeV]14.3 & 0.09 & 0.11 & 0.72 & 1.00 \\
 & {[ArVI]}4.53/[NeV]14.3 & 0.10 & 0.14 & 0.03 & 0.04 \\
 & {[NeVI]}7.68/[NeV]14.3 & 0.58 & 0.78 & 0.04 & 0.06 \\[0.15cm] \cline{3-6} 
\bf SFG & \multicolumn{1}{r|}{log\,U} & -2.5 & -2.5 & -3.5 & -3.5 \\
 & \multicolumn{1}{r|}{log(n$_{H}$/cm$^{-3}$)} & 1 & 3 & 1 & 3 \\[0.1cm] \cline{3-6} 
 & {[ArII]}6.98/[NeII]12.8 & 0.05 & 0.06 & 0.12 & 0.12 \\
 & {[ArIII]}8.99/[NeII]12.8 & 0.66 & 0.68 & 0.32 & 0.32 \\
 & $\frac{\rm [ArII]6.98 + [ArIII]8.99}{\rm [NeII]12.8 + [NeIII]15.6}$ & 0.30 & 0.30 & 0.38 & 0.38 \\[0.15cm] \cline{3-6} 
\bf LMG & \multicolumn{1}{r|}{log\,U} & -2 & -2 & -3 & -3 \\
 & \multicolumn{1}{r|}{log(n$_{H}$/cm$^{-3}$)} & 1 & 3 & 1 & 3 \\[0.1cm] \cline{3-6} 
 & {[ArII]}6.98/[NeII]12.8 & 0.12 & 0.12 & 0.12 & 0.11 \\
 & {[ArIII]}8.99/[NeII]12.8 & 8.31 & 8.29 & 1.41 & 1.40 \\
 & $\frac{\rm [ArII]6.98 + [ArIII]8.99}{\rm [NeII]12.8 + [NeIII]15.6}$ & 0.19 & 0.18 & 0.27 & 0.27 \\[0.15cm]
\hline
\end{tabular}%
}
\end{table}

\section{Methods}
\label{sec:methods}

We assess how the JWST-MIRI and ALMA instruments can measure the star formation rate (SFR) and the black hole accretion rate (BHAR) covering, respectively, the z$\lesssim$3 and z$\gtrsim$3 redshift intervals. The predictions for JWST-MIRI have been derived using the CLOUDY \citep{ferland2017} photo-ionization models computed by \citet{fernandez2016}, using the BPASS v2.2 stellar population synthetic library \citep{stanway2018} for the case of the star forming models. We consider, in particular, three classes of objects, as presented in \citet{mordini2021}: \textit{i)} AGN, \textit{ii)} star forming galaxies (SFG), and \textit{iii)} low metallicity galaxies (LMG) covering the $\sim$7$\leq$12 + log (O/H)$\leq$8.5 metallicity range \citep{madden2013,cormier2015}. This is motivated by the detection of massive galaxies ($\sim$10$^{10}$ M$_{\odot}$) in optical surveys with sub-solar metallicities during the Cosmic Noon, which may experience a fast chemical evolution above z$\sim$2 \citep{maiolino2008,mannucci2009,troncoso2014,onodera2016}. These changes are expected to have an impact on the ISM structure of high-z galaxies, favouring stronger radiation fields \citep[e.g.][]{steidel2016,kashino2019,sanders2020} and a more porous ISM similar to that observed in dwarf galaxies \citep{cormier2019}.

We have explored the well known problem of the relative weakness of the narrow lines intensities in Seyfert galaxies and quasars at increasing bolometric luminosity, which has been observed in optical forbidden lines \citep[e.g., ][]{stern2012}. We have tested this hypothesis on a sample of bright PG quasars  with [NeV]14.3$\mu$m  and [OIV]25.9$\mu$m measurements from \citet{veilleux2009}. For these high-excitation lines, the depression of the line intensities is noticeable for total IR luminosities above 10$^{45}~{\rm erg\,s^{-1}}$.  This value is larger than the luminosity range covered by the AGN sample used to derive the IR fine structure line calibrations ($10^{42}<{\rm L_{IR}}< 10^{45}~{\rm erg\,s^{-1}}$; \citealt{mordini2021}), thus we conclude that this effect is not biasing our results. It follows, however, that our conclusions are limited to AGN with total luminosities below this threshold, while for brighter objects the predicted accretion luminosities and the accretion rates could be underestimated.

To measure the BHAR, we consider the [MgIV] 4.49$\mu$m, [ArVI] 4.53$\mu$m \citep[see, e.g.,][]{satyapal2021} and the [NeVI] 7.65$\mu$m line. We need to caution the reader about the use of the [NeVI]7.65$\mu$m line, because its detections by the Infrared Space Observatory \citep[ISO;][]{sturm2002}, on which is based the calibration \citep{mordini2021} are scarce and have only been obtained for the brightest sources, and therefore this calibration is uncertain. 
The properties of these lines are summarized in Table \ref{tab:line_properties}.  

For the BHAR tracers, we use the template of an AGN of total IR luminosity of L$_{IR}$=10$^{12}$L$_{\odot}$, to represent the bulk of the population at redshift z$\sim$2. We use four different photo-ionization models to derive the line intensities, with two values of the ionization parameter: log\,$U$=-1.5 and log\,$U$=-2.5, and two values of the hydrogen density: log(n$_{H}$/cm$^{-3}$)=2 and log(n$_{H}$/cm$^{-3}$)=4. The calibration of the line luminosities, against the total IR luminosities, has been derived using the ratios of the high-excitation magnesium, argon and neon lines relative to the [NeV]14.3$\mu$m line in the photo-ionization models, and the calibration of the [NeV]14.3$\mu$m line from \citet{mordini2021}.

Table \ref{tab:line_properties} gives the key properties of the lines used in this work. Table \ref{tab:line_ratios} reports the line ratios used to derive the line intensities for the [MgIV]4.49$\mu$m, [ArVI]4.53$\mu$m, [ArII]6.98$\mu$m, [NeVI]7.65$\mu$m and [ArIII]8.99$\mu$m lines, given the different combinations of ionization parameter and hydrogen densities. Table \ref{tab:derived_correlations} reports the calibrations computed from the photo-ionization models scaled to the [NeII]12.8 $\mu$m and [NeV]14.3$\mu$m lines (see Section \ref{sec:results_jwst}), while the calibrations of the different lines and features in the mid- to far-IR range can be found in Table D.1 in \citet{mordini2021}.

The line ratios derived from the [NeVI] and [MgIV] detections reported in \citet{sturm2002} are all consistent with AGN characterized by a ionization parameter of log\,$U$=-1.5 and hydrogen density of log(n$_{H}$/cm$^{-3}$)=4 (see Table \ref{tab:line_ratios}), or higher, with line ratios of  [NeVI]/[NeV] $\sim$0.67-1.4 and [MgIV]/[NeV] $\sim$0.06. The low number of detections (8 for the [NeVI] line and 3 for the [MgIV] line) suggests that while the [NeVI] line is more easily observable in AGN whose ISM is characterized by a high ionization potential, the average AGN is relatively weaker. For this reason, while the [MgIV] line can give an accurate measure of the BHAR in average AGN, a comprehensive measure of AGN activity can be achieved using a combination of [MgIV] and [ArVI] or [NeVI] lines, in order to include also powerful AGN.

As SFR tracers, we take into account the [ArII]6.98$\mu$m and [ArIII]8.99$\mu$m lines. We consider both SFG and LMG, deriving, in both cases, the line intensities dependent on the calibration of the [NeII]12.8$\mu$m line \citep{mordini2021} considering, for the SFG, photo-ionization models with an ionization parameter of log\,$U$=-2.5 and log\,$U$=-3.5, and a hydrogen density of log(n$_{H}$/cm$^{-3}$)=1 and log(n$_{H}$/cm$^{-3}$)=3, for a total of four different cases. For the LMG we consider an ionization parameter of log\,$U$=-2 and log\,$U$=-3, and hydrogen densities of log(n$_{H}$/cm$^{-3}$)=1 and log(n$_{H}$/cm$^{-3}$)=3. In both cases, we assume a total IR luminosity of the galaxy of L$_{IR}$=10$^{12}$L$_{\odot}$.

The predictions for the ALMA telescope were derived from the line calibrations described in \citet{mordini2021}. While the spectral range covered by ALMA has no unambiguous high ionization line that can be used to trace the BHAR, we explore what lines, or combination of them, can be used to trace the SFR.
Following \citet{mordini2021}, we consider the [CII]158$\mu$m line and the sum of two oxygen lines, either the [OI]63$\mu$m and [OIII]88$\mu$m, or the [OI]145$\mu$m and [OIII]88$\mu$m.

\begin{table}
\caption{{ Correlations} of the fine-structure line luminosities { with} the total IR luminosities (log\,$L_{\rm Line}$\,=\,$a$\,log\,$L_{IR}$\,+\,$b$), derived from the line ratios reported in Table \ref{tab:line_ratios}. For AGN: log\,$U$\,=\,-2.5 and log(n$_{H}$/cm$^{-3}$)\,=\,2; for SFG: log\,$U$\,=\,-3.5 and  log(n$_{H}$/cm$^{-3}$)\,=\,3  and for LMG: log\,$U$\,=\,-2 and log(n$_{H}$/cm$^{-3}$)\,=\,1 .}
\label{tab:derived_correlations}
\resizebox{\textwidth}{!}{%
\begin{tabular}{l|cc|cc|cc}
\hline
\bf Line & \multicolumn{2}{c|}{\bf AGN} & \multicolumn{2}{c|}{\bf SFG} & \multicolumn{2}{c}{\bf LMG} \\
 & $a$ & $b$ & $a$ & $b$  & $a$  & $b$  \\ \hline
{[MgIV]}4.49 & 1.32 & -5.15 & -- & -- & -- & -- \\
{[ArVI]}4.53 & 1.32 & -6.53 & -- & -- & -- & -- \\
{[NeVI]}7.65 & 1.32 & -6.41 & -- & -- & -- & -- \\
{[ArII]}6.98 & -- & -- & 1.04 & -4.4 & 1.37 & -5.28 \\
{[ArIII]}8.99 & -- & -- & 1.04 & -3.97 & 1.37 & -3.44
\end{tabular}%
}
\end{table}

\section{Results}
\label{sec:results}

\subsection{Predictions for JWST}
\label{sec:results_jwst}
 
\begin{figure}[!ht]
    \centering
    \includegraphics[width=0.95\textwidth]{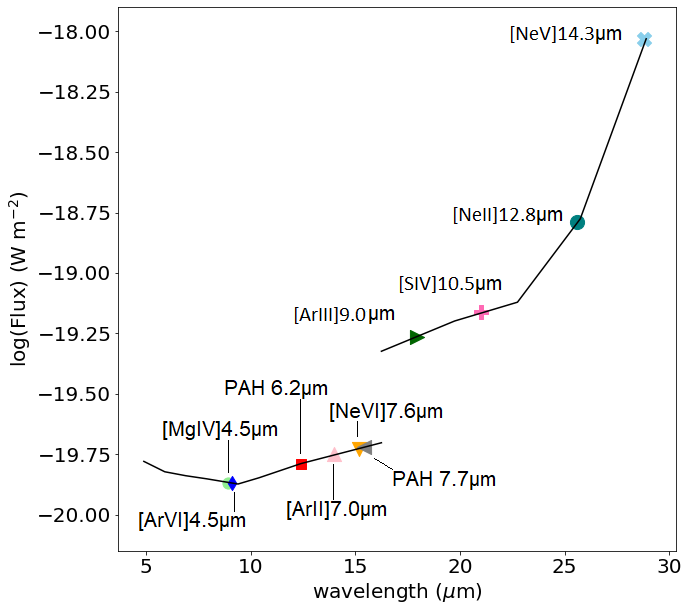}
    \caption{Sensitivity of the JWST-MIRI instrument at different wavelengths, for a 1 hr., 5$\sigma$ observation. The symbols show the position, in the MIRI wavelength range, at which the different lines are observed in a galaxy at redshift z = 1. From left to right, the considered lines and feature, at their rest frame wavelength, are: [MgIV]4.49$\mu$m, [ArVI]4.53$\mu$m, PAH feature at 6.2$\mu$m, [ArII]6.98$\mu$m, [NeVI]7.65$\mu$m, PAH feature at 7.7$\mu$m, [ArIII]8.99$\mu$m, [SIV]10.5$\mu$m, [NeII]12.8$\mu$m and [NeV]14.3$\mu$m.}
    \label{fig:miri_z1}
\end{figure}

In this section, we predict the intensities of the mid-IR lines and features in the 5-9$\mu$m { rest-frame} range, as derived from CLOUDY models, and compare them with the JWST-MIRI sensitivity, to verify if these lines, used as tracers for SFR and BHAR, can be observed in { galaxies and AGN} at redshifts z$>$1. Table \ref{tab:derived_correlations} reports the derived correlations of the lines used in this work for the most common values of ionization potential (log\,$U$) and gas density [log(n$_{H}$/cm$^{-3}$)] for each type of object. For the AGN lines, we use the calibration of the [NeV]14.3$\mu$m line, while for SFG and LMG we use the [NeII]12.8$\mu$m line. Since these correlations are derived from CLOUDY simulations, we do not report the error on the coefficients. Fig.\,\ref{fig:miri_z1} shows the sensitivity of JWST-MIRI, highlighting the wavelengths at which the various fine-structure lines and PAH spectral features can be observed in a galaxy at redshift z=1. In the $\sim 5-16\mu$m { observed spectral range}, the MIRI sensitivity is below $\sim$2$\times$10$^{-20}$W\,m$^{-2}$ for a 1hr, 5$\sigma$ observation, while in the $\sim 16-29\mu$m { observed spectral range} the sensitivity is increasingly worse, going from $\sim$6$\times$10$^{-20}$W\,m$^{-2}$ at $\sim$16$\mu$m, to $\sim$10$^{-18}$W\,m$^{-2}$ at $\sim$29$\mu$m. This makes observing the [SIV]10.5$\mu$m, [NeII]12.8$\mu$m and [NeV]14.3$\mu$m lines { at redshift z=1} a challenging task, and rises the need to find and analyze { other} tracers.

In the 5-16$\mu$m { observed spectral} range,  { AGN tracers} are the [MgIV]4.49$\mu$m, [ArVI]4.53$\mu$m and [NeVI]7.65$\mu$m lines, while the [ArII]6.98$\mu$m and the [ArIII] 8.99$\mu$m lines { are} tracers for star formation activity, together with the PAH feature at 6.2$\mu$m. Given their relatively short wavelength, these lines can be used to probe the highly obscured galaxies at the Cosmic Noon (1$\lesssim$z$\lesssim$3). While lines in the mid- to far-IR range do not typically suffer from significant dust extinction, the [MgIV] and [ArVI] lines, at wavelengths of $\lambda \sim 4.5$, can be affected by the presence of heavy dust obscuration. In particular, an optical extinction of $A_{V}\sim 10\, \rm{mag}$ would absorb about $\sim 30\%$ of the radiation at 4.5$\mu$m, following the extinction curve of \citet{cardelli1989}. This value increases to $\sim 80\%$ for { the very high optical extinction} of $A_{V}\sim 50\, \rm{mag}$.

\begin{figure*}
    \centering
    \includegraphics[width=0.45\textwidth]{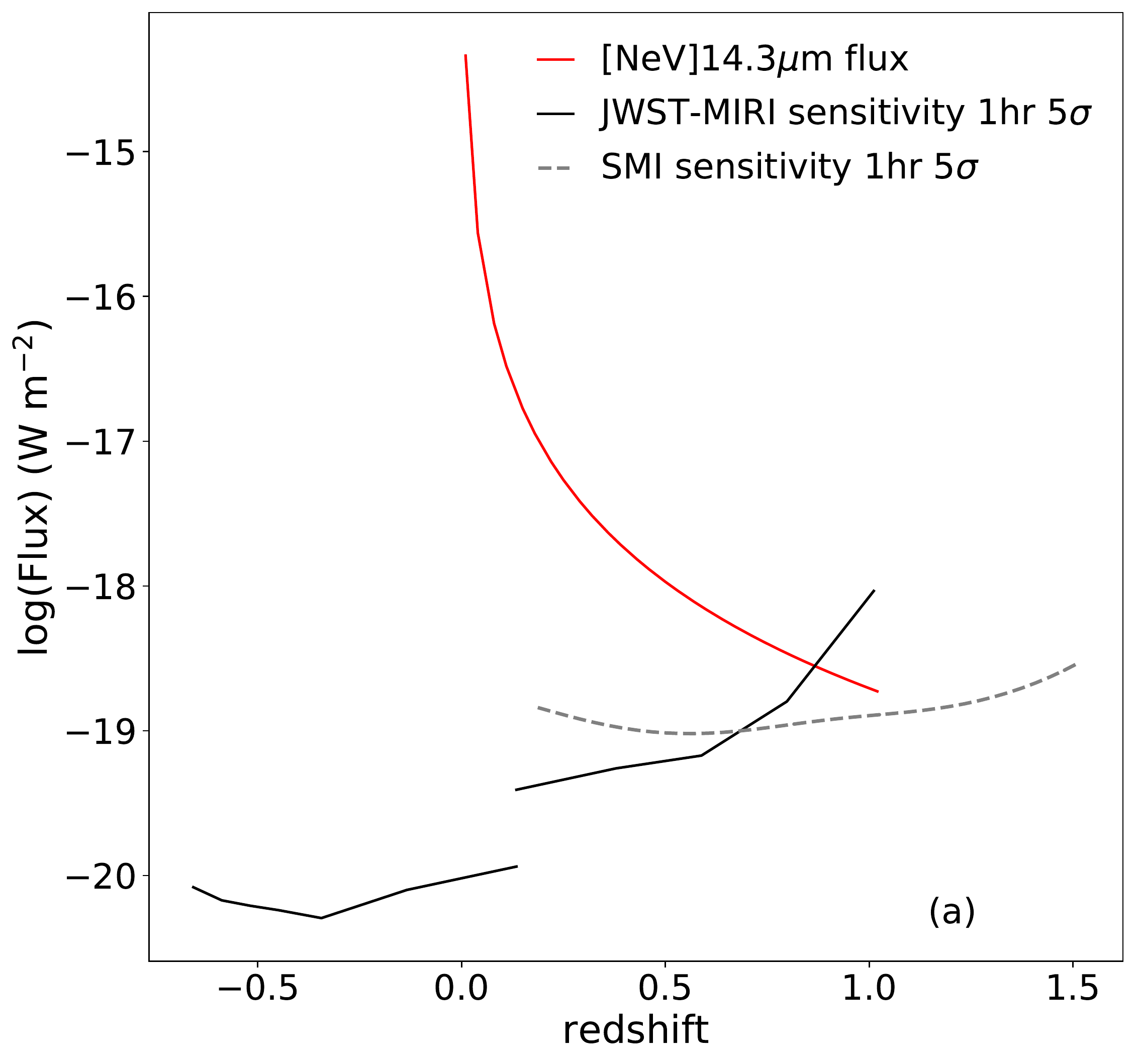}~
    \includegraphics[width=0.45\textwidth]{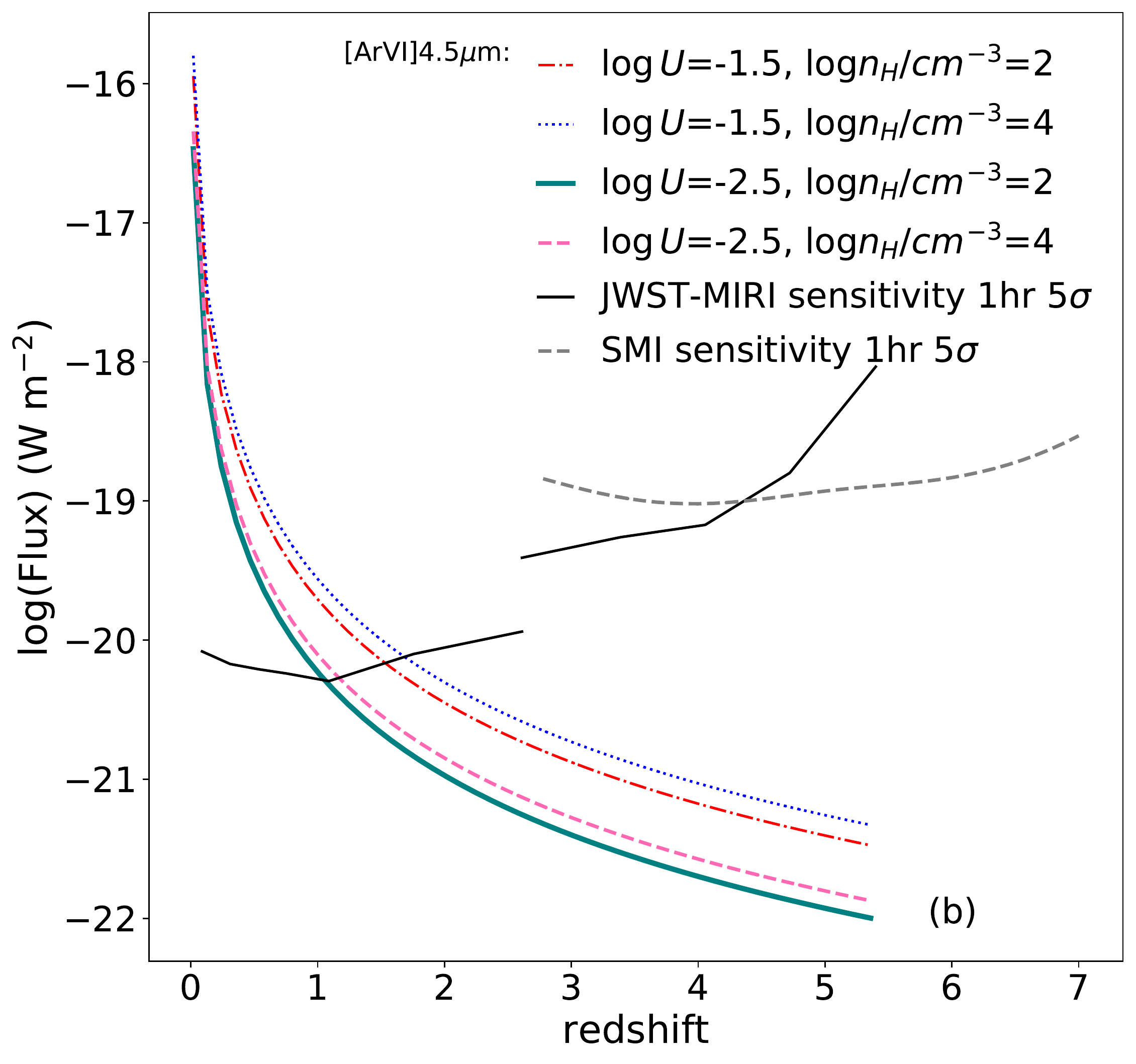}
    \includegraphics[width=0.45\textwidth]{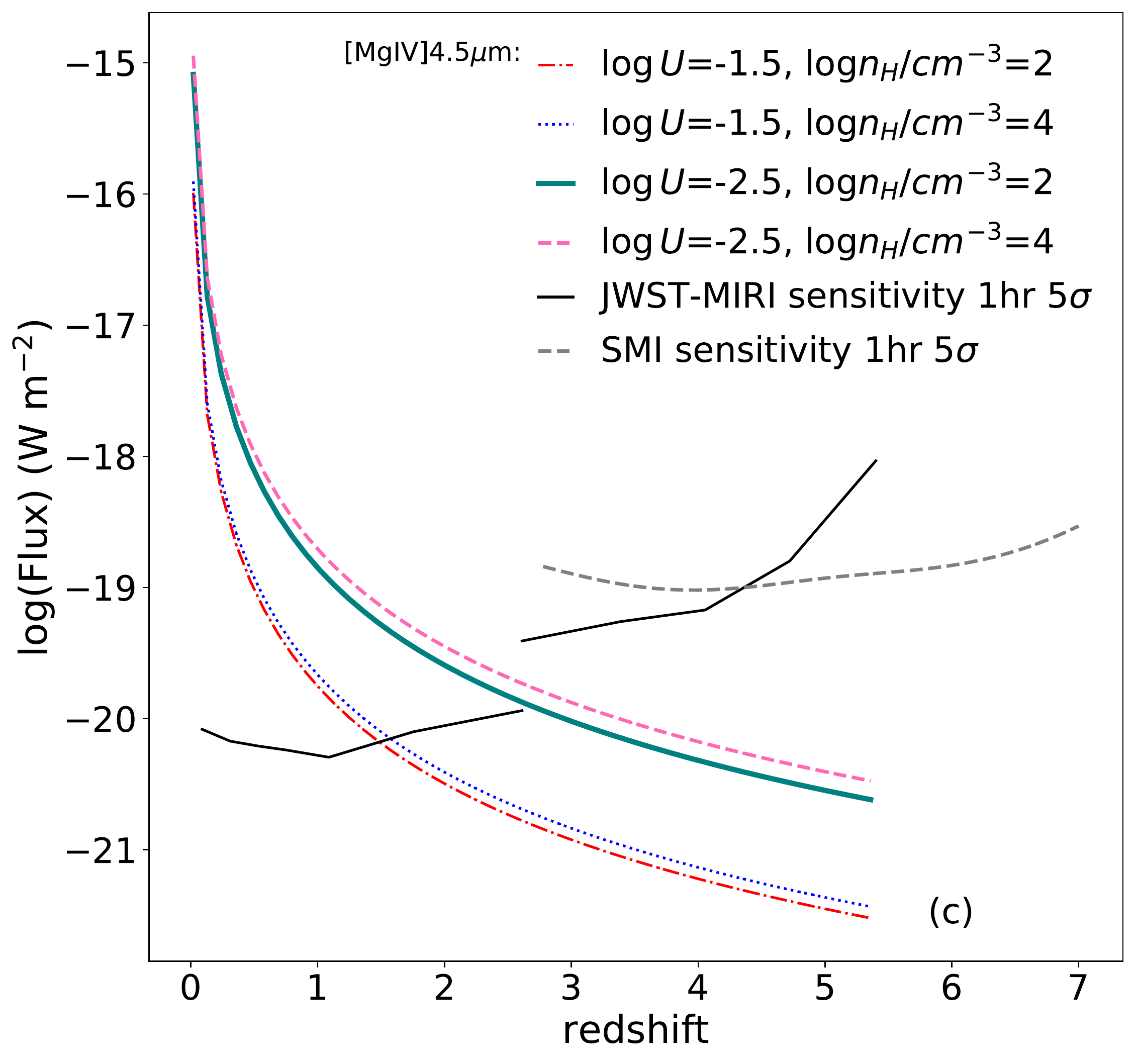}~
    \includegraphics[width=0.45\textwidth]{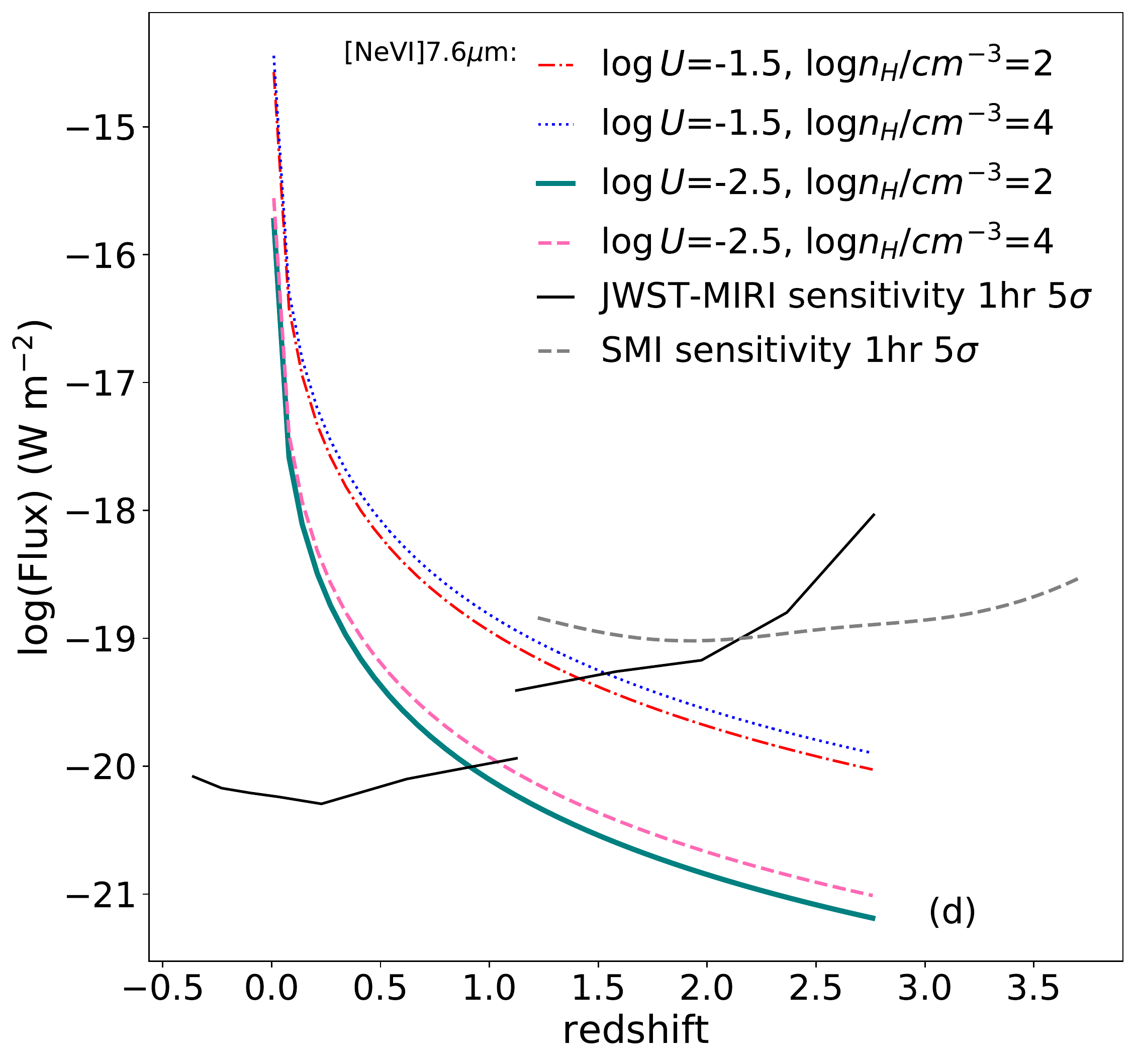}
    \caption{Predicted fluxes, as a function of redshift, for an AGN with a  total IR luminosity of L$_{IR}$=10$^{12}$L$_{\odot}$ for the [NeV]14.3$\mu$m  (a:top left, red solid line); the [ArVI]4.5$\mu$m line (b:top right); the [MgIV]4.49$\mu$m line {\bf (c:bottom left)} and  for the [NeVI]7.65$\mu$m line (d: bottom right). In all figures, the black solid line shows the 1 hr., 5 $\sigma$ sensitivity of JWST-MIRI, while the grey dashed line shows the 1 hr., 5 $\sigma$ sensitivity of the SPICA SMI-LR \citep{kaneda2017}. In panels b, c and d, the red dash-dotted line shows the predicted flux for a galaxy with an ionization parameter of log\,$U$=-1.5 and a hydrogen density of log(n$_{H}$/cm$^{-3}$)=2, the blue dotted line indicates log\,$U$=-1.5 and log(n$_{H}$/cm$^{-3}$)=4, the green solid line shows log\,$U$=-2.5 and log(n$_{H}$/cm$^{-3}$)=2, and the pink dashed line shows log\,$U$=-2.5 and log(n$_{H}$/cm$^{-3}$)=4.}
    \label{fig:miri_ar6_mg4_ne6}
\end{figure*}

\begin{figure*}
    \centering
    \includegraphics[width=0.45\textwidth]{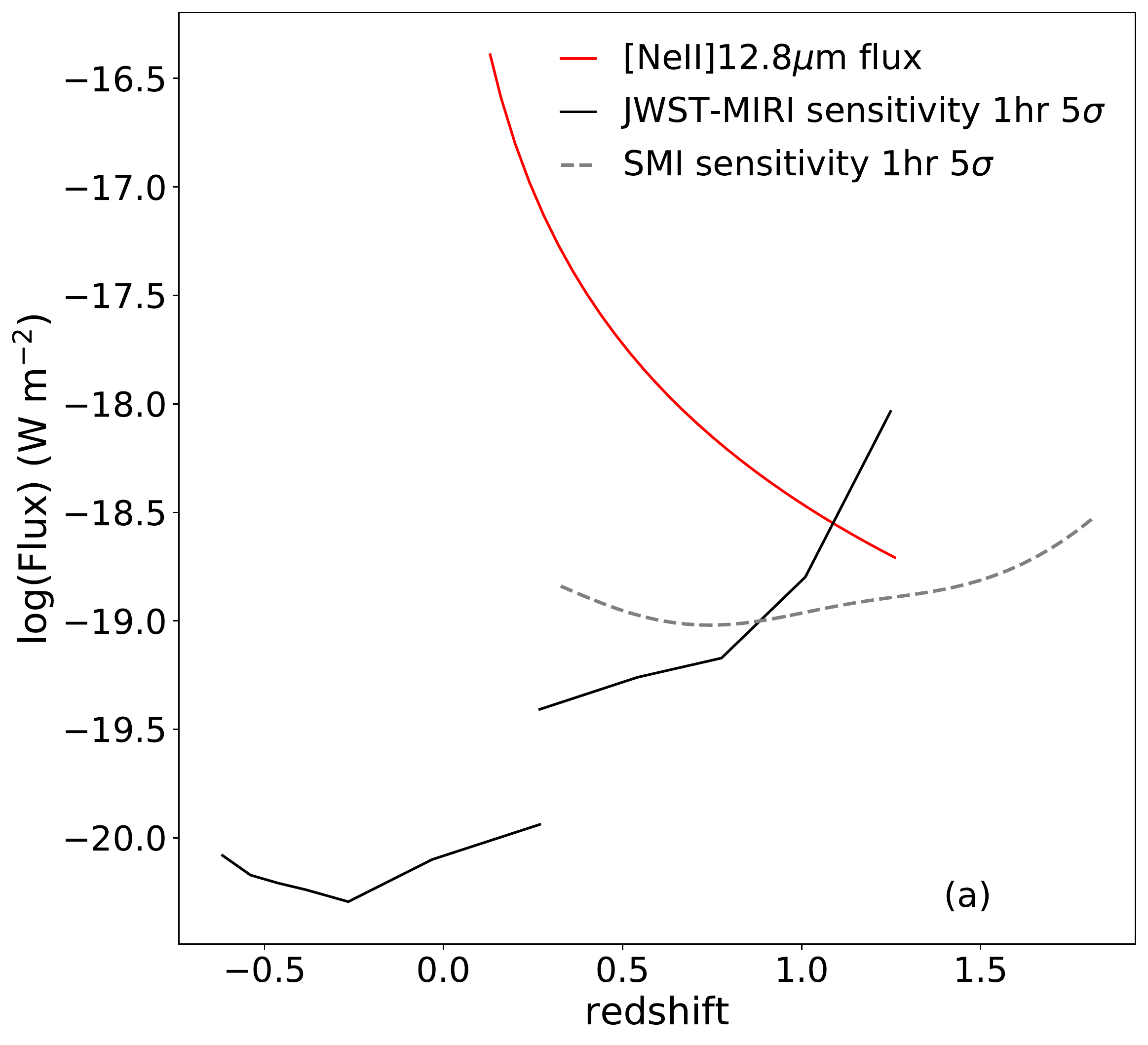}~
    \includegraphics[width=0.45\textwidth]{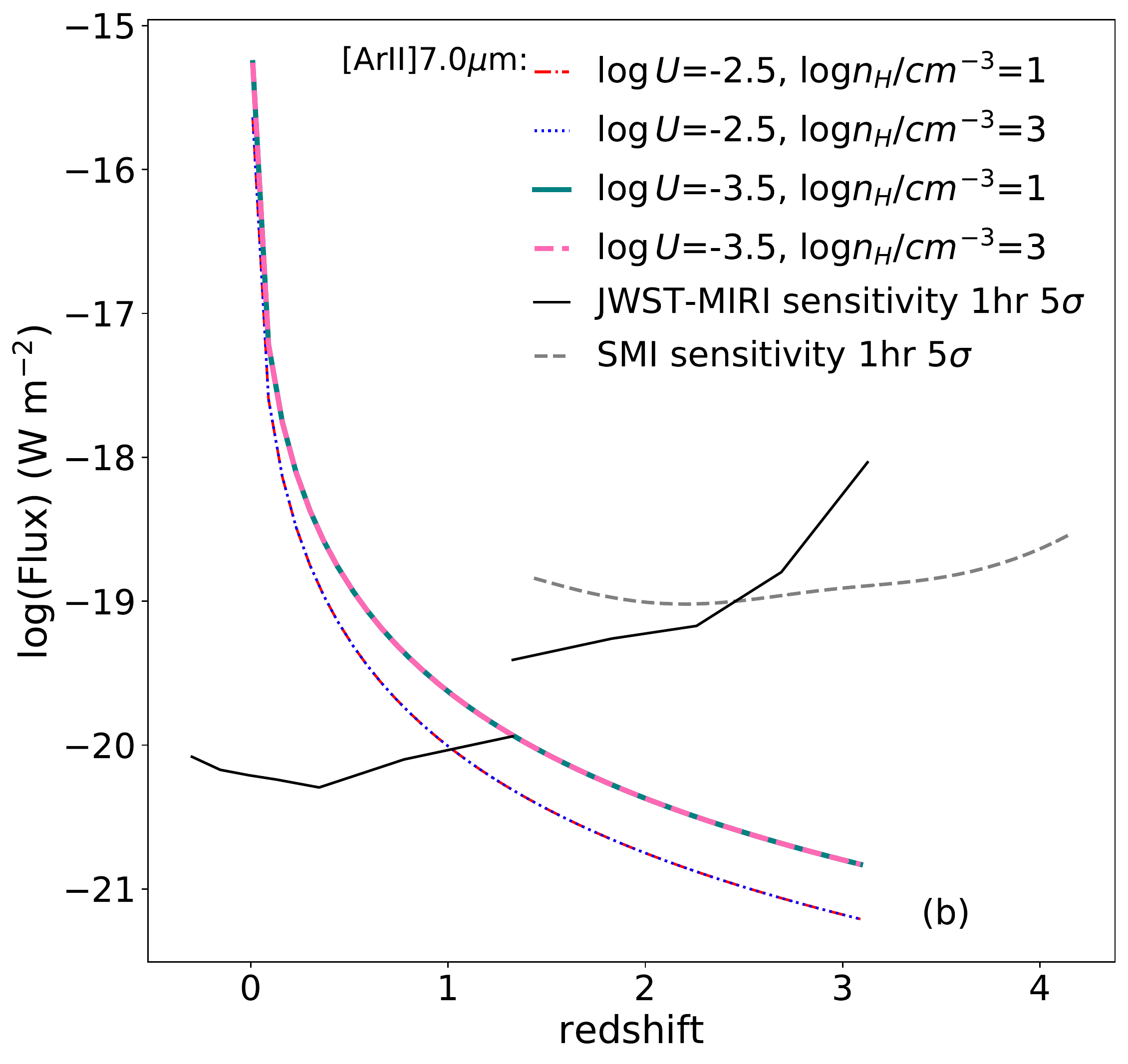}
    \includegraphics[width=0.45\textwidth]{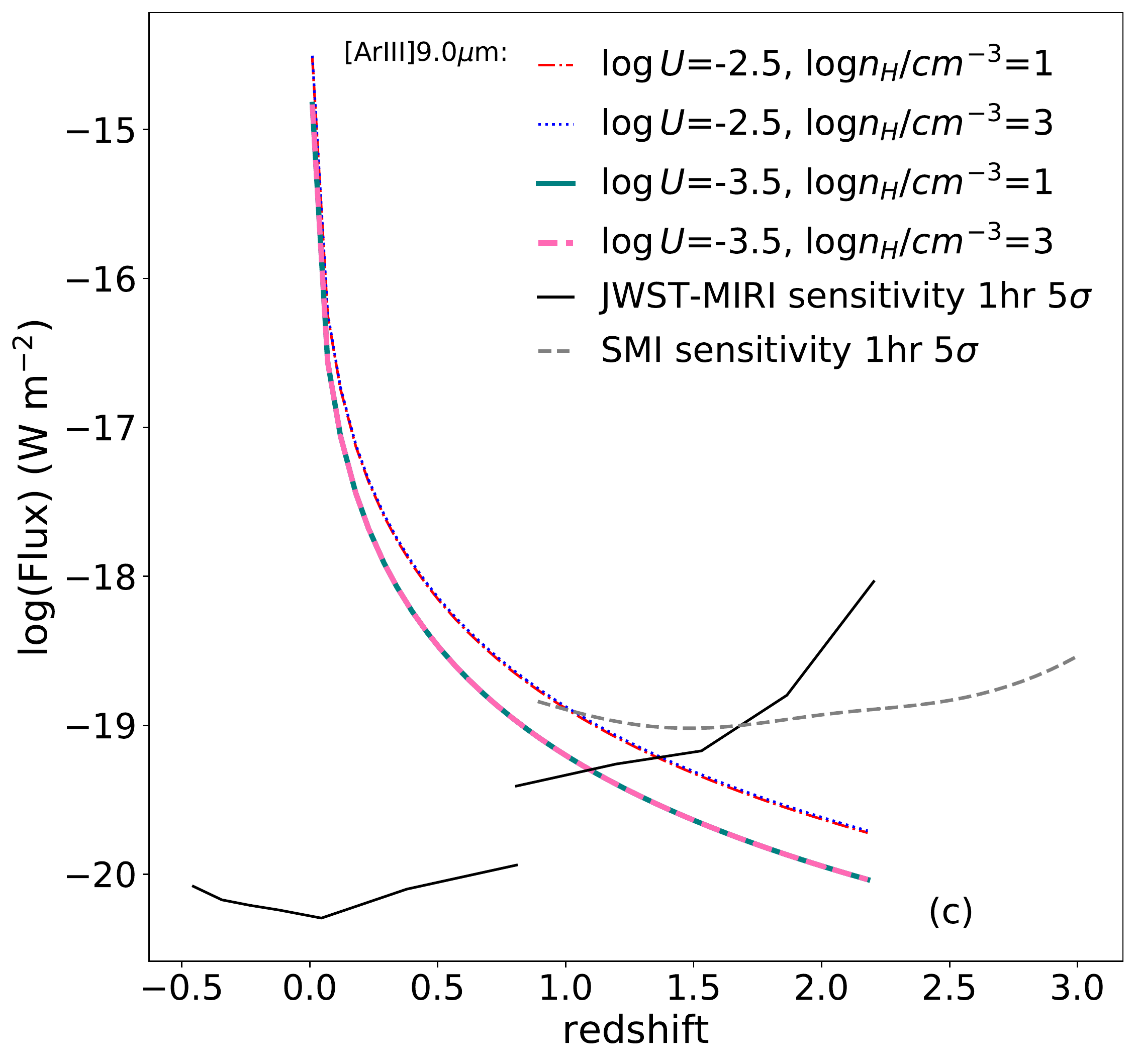}
    \caption{Predicted fluxes, as a function of redshift, of a SFG with a total IR luminosity of L$_{IR}$=10$^{12}$L$_{\odot}$ for the [NeII]12.8$\mu$m  (a:top left, red solid line); [ArII]6.98$\mu$m line (b:top right) and for the [ArIII]8.99$\mu$m line (c: bottom). In all figures, the black solid line shows the 1 hr., 5 $\sigma$ sensitivity of JWST-MIRI, while the grey dashed line shows the 1 hr., 5 $\sigma$ sensitivity of the SPICA SMI-LR \citep{kaneda2017}. In panels b and c, the red dash-dotted line shows a galaxy with an ionization parameter of log\,$U$=-2.5 and a hydrogen density of log(n$_{H}$/cm$^{-3}$)=1, the blue dotted line indicates log\,$U$=-2.5 and log(n$_{H}$/cm$^{-3}$)=3, the green solid line shows log\,$U$=-3.5 and log(n$_{H}$/cm$^{-3}$)=1, and the pink dashed line shows log\,$U$=-3.5 and log(n$_{H}$/cm$^{-3}$)=3.}
    \label{fig:miri_ar2_ar3_sf}
\end{figure*}

\begin{figure*}
    \centering
    \includegraphics[width=0.45\textwidth]{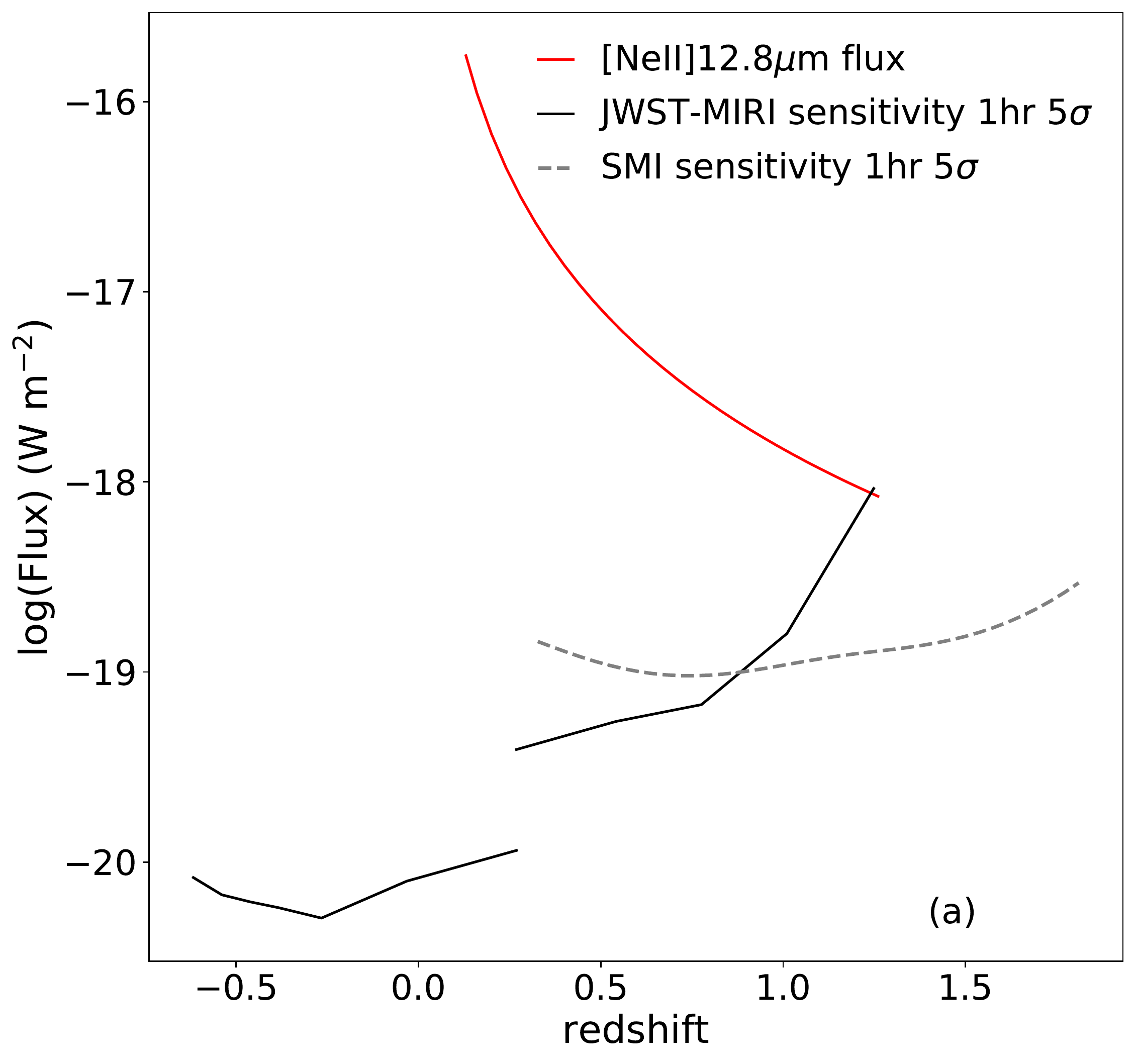}~
    \includegraphics[width=0.45\textwidth]{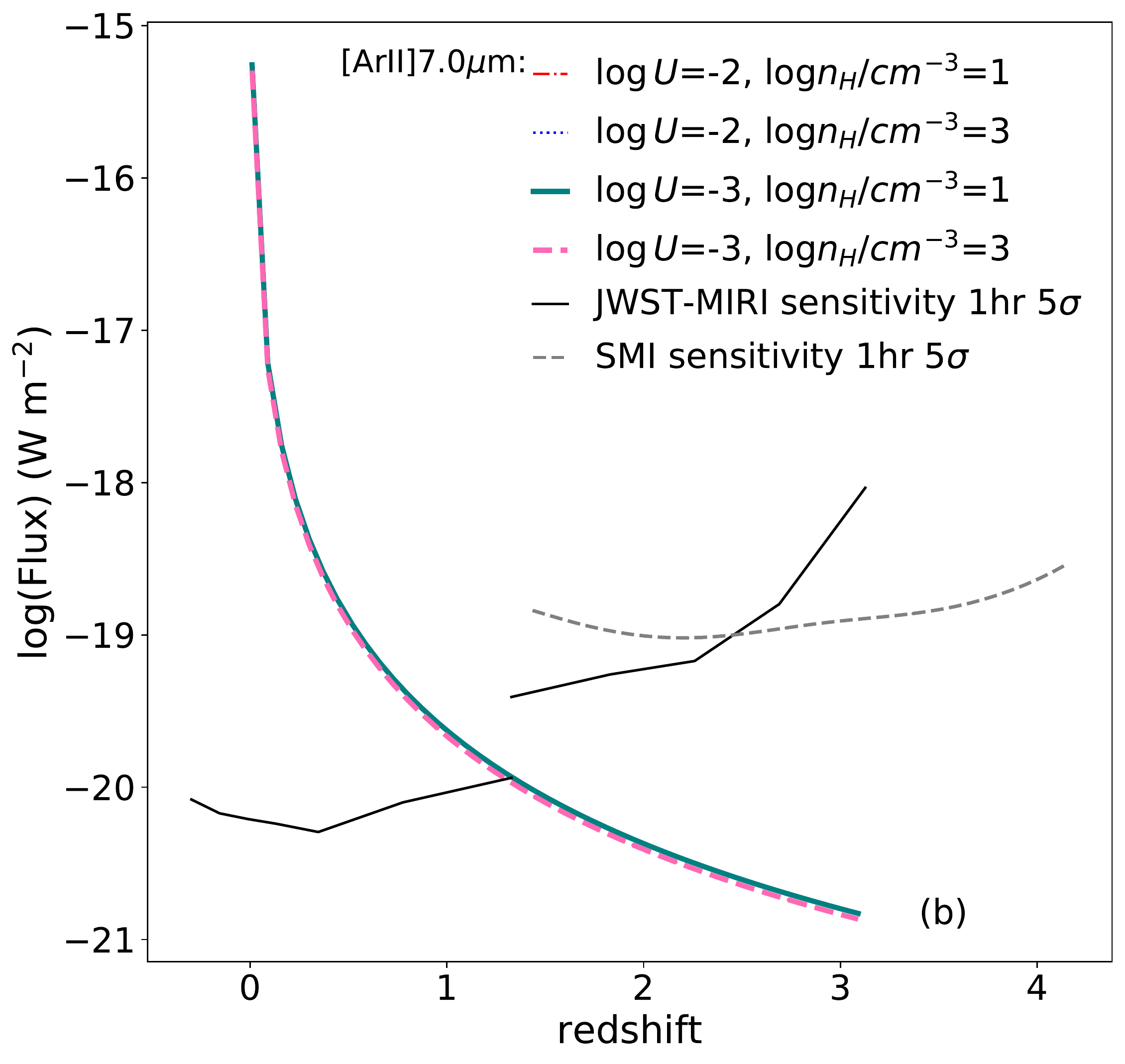}
    \includegraphics[width=0.45\textwidth]{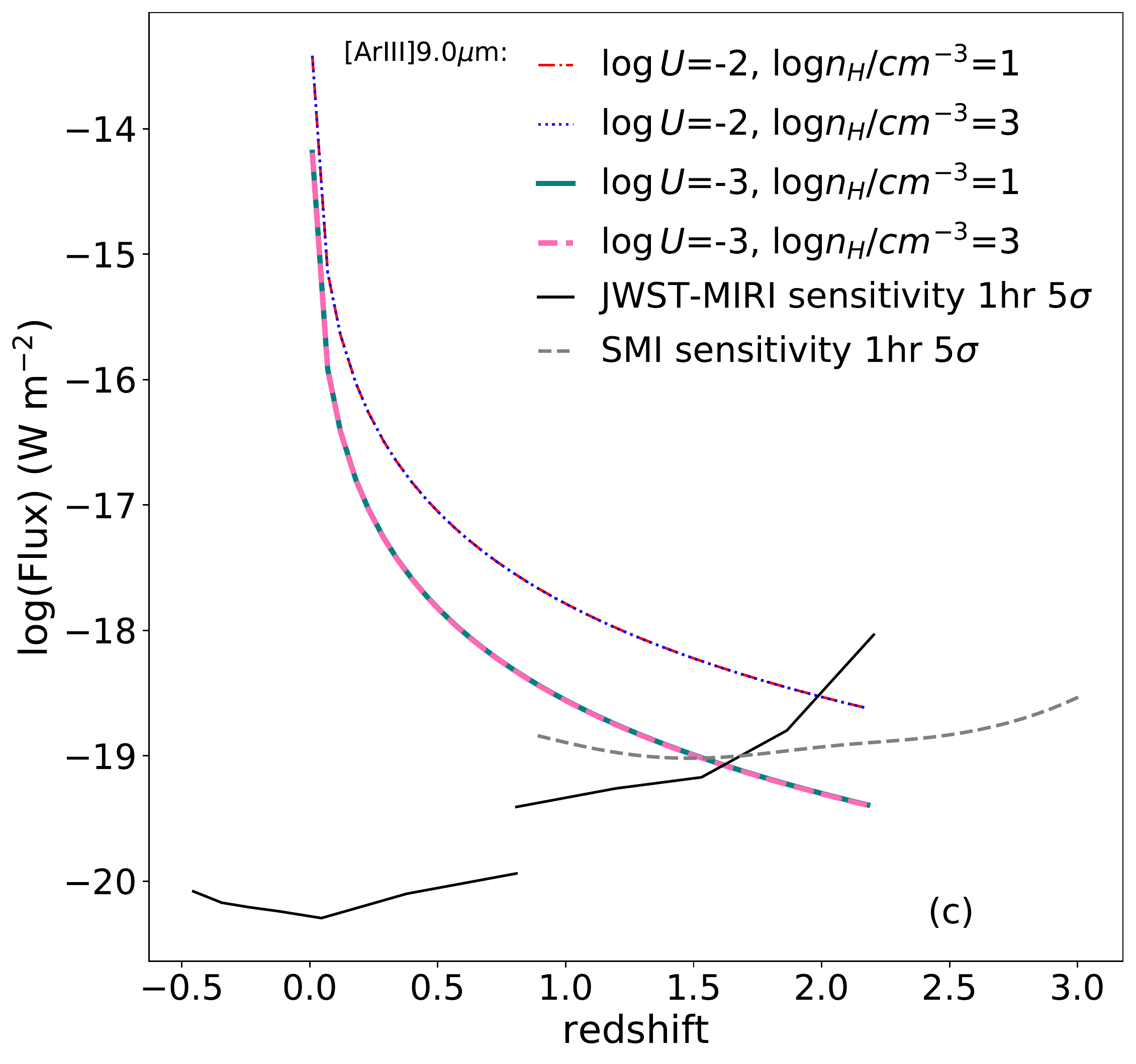}
    \caption{Predicted fluxes, as a function of redshift, of a LMG with a total IR luminosity of L$_{IR}$=10$^{12}$L$_{\odot}$ for the [NeII]12.8$\mu$m (a:top left, red solid line); the [ArII]6.98$\mu$m line (b:top right) and the [ArIII]8.99$\mu$m line (c: bottom). In all figures, the black solid line shows the 1 hr., 5 $\sigma$ sensitivity of JWST-MIRI, while the grey dashed line shows the 1 hr., 5 $\sigma$ sensitivity of the SPICA SMI-LR \citep{kaneda2017}. In panels b and c, the red dash-dotted line shows a galaxy with an ionization parameter of log\,$U$=-2 and a hydrogen density of log(n$_{H}$/cm$^{-3}$)=1, the blue dotted line indicate log\,$U$=-2 and log(n$_{H}$/cm$^{-3}$)=3, the green solid line shows log\,$U$=-3 and log(n$_{H}$/cm$^{-3}$)=1, and the pink dashed line shows log\,$U$=-3 and log(n$_{H}$/cm$^{-3}$)=3.}
    \label{fig:miri_ar2_ar3_lmg}
\end{figure*}

\begin{figure*}
    \centering
    \includegraphics[width=0.45\textwidth]{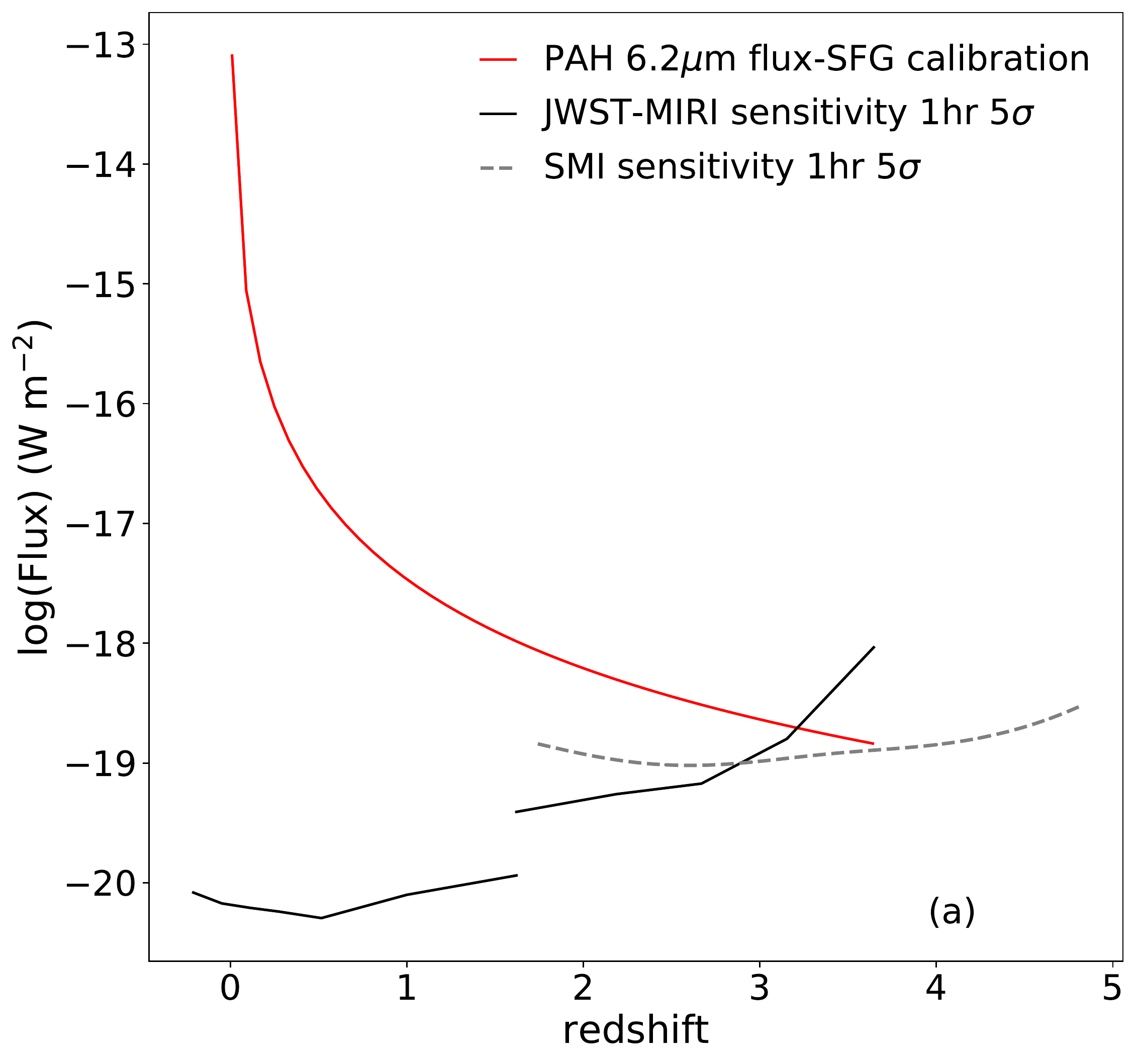}~
    \includegraphics[width=0.45\textwidth]{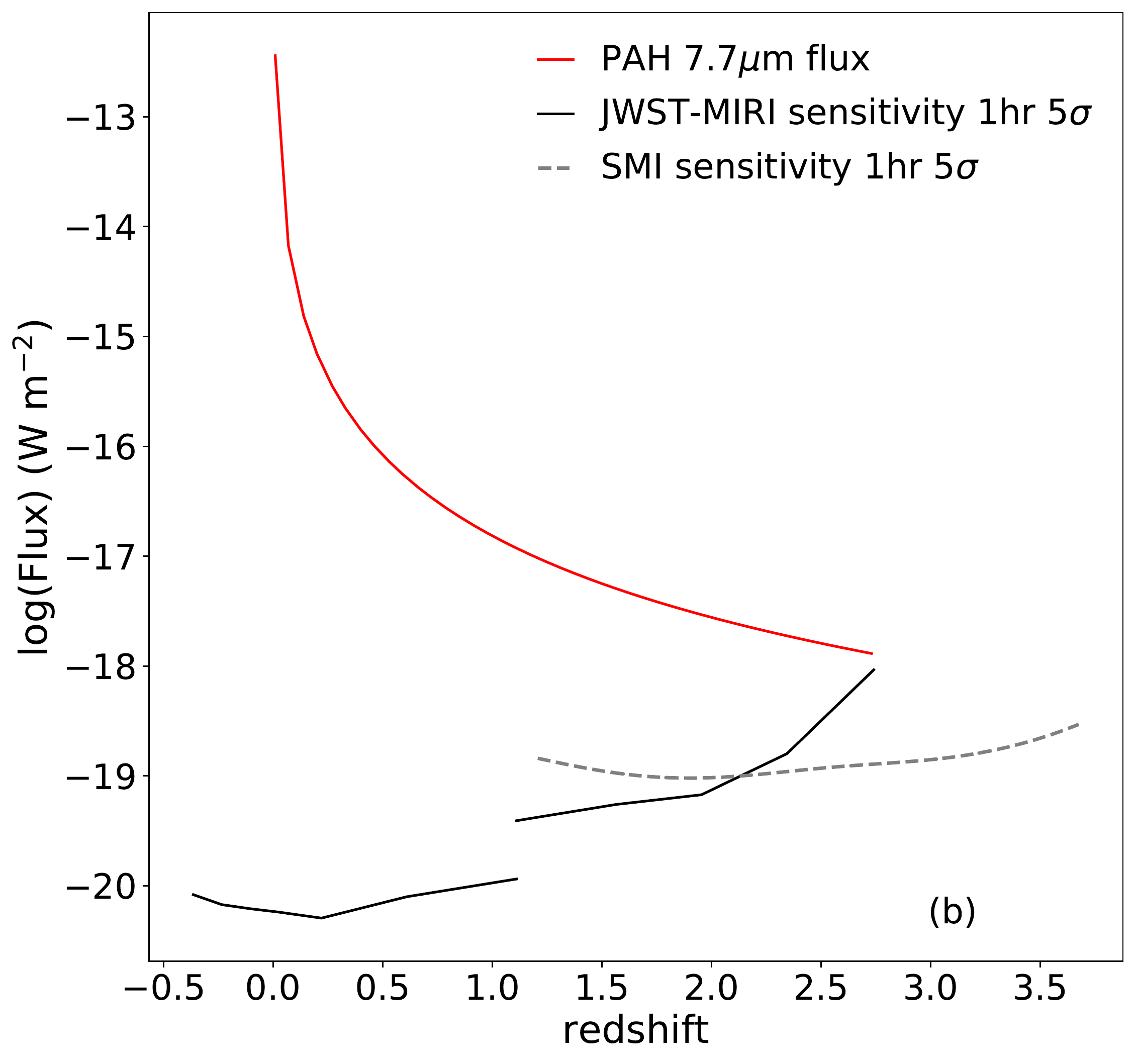}
    \includegraphics[width=0.45\textwidth]{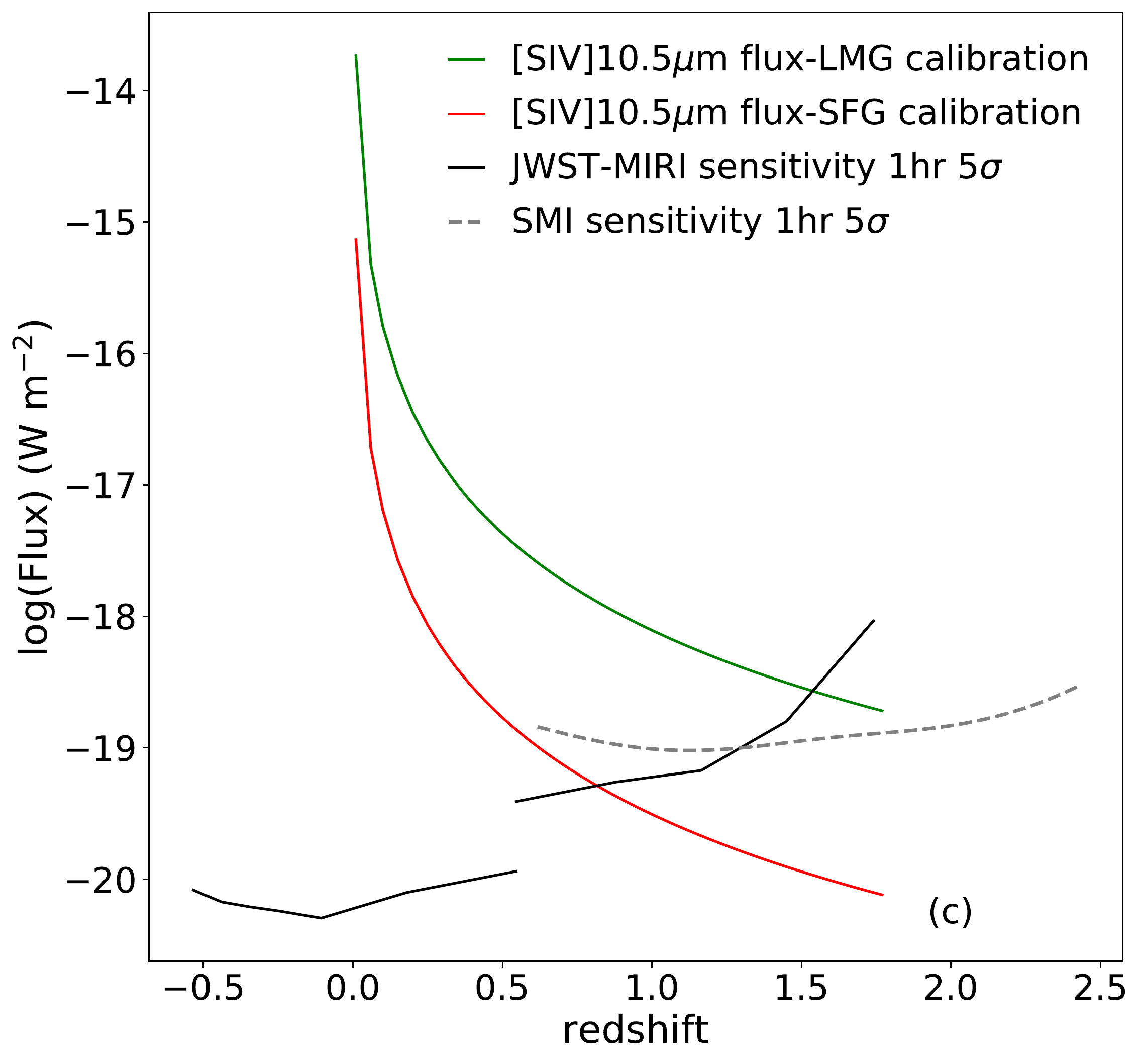}
    \caption{Predicted fluxes, as a function of redshift, of a source with a total IR luminosity of L$_{IR}$=10$^{12}$L$_{\odot}$ for the 6.2$\mu$m PAH feature (a:top left); the 7.7$\mu$m PAH feature (b:top right) and for the [SIV]10.5$\mu$m line considering the SFG calibration (red solid line) and the LMG calibration (green solid line) (c:bottom). In all figures, the black solid line shows the 1 hr., 5 $\sigma$ sensitivity of JWST-MIRI, while the grey dashed line shows the 1 hr., 5 $\sigma$ sensitivity of the SPICA SMI-LR \citep{kaneda2017}.}
    \label{fig:miri_pah_s4}
\end{figure*}

Fig.\,\ref{fig:miri_ar6_mg4_ne6} shows the predicted fluxes, as a function of redshift, compared to the MIRI sensitivity for the [NeV]14.3$\mu$m (panel a), [ArVI]4.53$\mu$m (panel b), [MgIV]4.49$\mu$m (panel c) and [NeVI]7.65$\mu$m (panel d) lines, for an AGN with a total IR luminosity of L$_{IR}$=10$^{12}$L$_{\odot}$. We chose this luminosity because it better represents the population at the knee of the luminosity function for galaxies at redshift z$\lesssim$3. For comparison, this figure and the following ones include also the sensitivity of the SPICA Mid-IR Instrument (SMI; grey dashed line) that was considered for the SPICA Mission \citep{roelfsema2018,spinoglio2021,kaneda2017}. This comparison indicates the better sensitivity of a cryogenically cooled space telescope at long wavelengths ($\lambda$>15$\mu$m), even of smaller diameter, such as the one of the SPICA project (2.5m), as compared to the JWST. 

Considering an integration time of 1~hr. and a signal to noise ratio of SNR=5, MIRI will be able to detect the [NeV] line up to redshift z$\sim$0.8. The [ArVI] line can be observed up to redshift z$\sim$1.8 for objects with an ionization parameter of $\log$U=-1.5, almost independently of the gas density, while objects with lower ionization can be observed up to redshift z$\sim$1. The [MgIV] line can be observed up to redshift z$\sim$1.8 for objects with an ionization parameter of $\log$U=-1.5, independently of the gas density. An ionization parameter of $\log$U=-2.5 { extends} the observational limit to { a redshift of} z$\sim$2.8.
The [NeVI]7.65$\mu$m line can be observed up to redshift z$\sim$1.2 for objects with an ionization parameter of $\log$U=-2.5, independently of gas density. However, for a higher ionization parameter of $\log$U=-1.5 and a gas density of $\log($n$_{H}$/cm$^{-3})$=2 this line can be detected up to redshift z$\sim$1.5, and up to redshift z$\sim$1.6 if the gas density is $\log($n$_{H}$/cm$^{-3})$=4.

\citet{satyapal2021}, performing similar simulations, assume different physical parameters. In particular, { these} authors fix { the hydrogen density to the value} of log(n$_{H}$/cm$^{-3}$) = 2.5, and assume a { ionization} parameter either equal to log\,$U$=-1 or -3. We obtain, however, similar results, with the [MgIV] line { which results to be} the strongest tracer in low ionization AGN, { while} the [NeVI] line { dominates the AGN spectra} at higher ionizations.

Fig.\,\ref{fig:miri_ar2_ar3_sf} and \ref{fig:miri_ar2_ar3_lmg} show the predicted fluxes { for a SFG and a LMG, respectively,} { with} total IR luminosity of L$_{IR}$=10$^{12}$L$_{\odot}$, as a function of redshift, compared to the MIRI sensitivity, of the [NeII]12.8$\mu$m  (panel a),  [ArII]6.98$\mu$m (panel b) and [ArIII]8.99$\mu$m (panel c) lines. For the SFG, the [NeII] line can be detected up to redshift z$\sim$1 with a 1 hr. observation, while the [ArII] line can be observed up to redshift z$\sim$1.5 for models with an ionization parameter of $\log$U=-2.5, while for $\log$U=-3.5 the limiting redshift is increased to z$\sim$2.2. For the [ArIII] line, the limiting redshift is of z$\sim$1.2 independently of the adopted model.

For the LMG, the simulation yields similar results. Considering a  source of total IR luminosity of L$_{IR}$=10$^{12}$L$_{\odot}$, the [NeII] line can be detected up to z$\sim$1.25; the [ArII] line can be detected up to z$\sim$1.5 independently of the adopted model. For the [ArIII] line the maximum redshift is reached for an ionization parameter of log\,$U$=-2, and is equal to z$\sim$2, while for an ionization parameter of log\,$U$=-3 the maximum redshift is of z$\sim$1.7, independent of gas density. 

Fig. \ref{fig:miri_pah_s4} shows the predicted fluxes { for both a SFG and LMG with total IR luminosity L$_{IR}$=10$^{12}$L$_{\odot}$}, as a function of redshift, compared to the MIRI sensitivity, { of} the PAH features at 6.2 and 7.7$\mu$m (panel a and b respectively), and { of} the [SIV]10.5$\mu$m line (panel c). Given the high intensity of the PAH features, these can be easily detected by MIRI up to redshift z$\sim$3.6 for the 6.2$\mu$m feature, and up to redshift z$\sim$2.7 for the 7.7$\mu$m feature. The [SIV] line can be observed up to redshift z$\sim$1.7, requiring longer integration times to be detected in SFG above redshift z$\sim$0.8. 

In all simulations, the decrease in sensitivity of MIRI in the 16-29$\mu$m { observed spectral range increases} the required exposure times to values much longer than 1 hr., in order to detect sources at higher redshift and thus probe the highly obscured galaxies at the Cosmic noon.

\subsection{Predictions for ALMA}
\label{sec:results_alma}

In this section we compare the predicted flux of the [OI]63 and 145$\mu$m, [OIII]88$\mu$m, [NII]122 and 205$\mu$m and [CII]158$\mu$m lines as a function of redshift to the sensitivity that the ALMA telescope can achieve in 1 hr. and 5 hr. observations, in order to determine the limits of the instrument in studying galaxy evolution through cosmic time. 

\begin{figure}
\centering
    \includegraphics[width=0.95\textwidth]{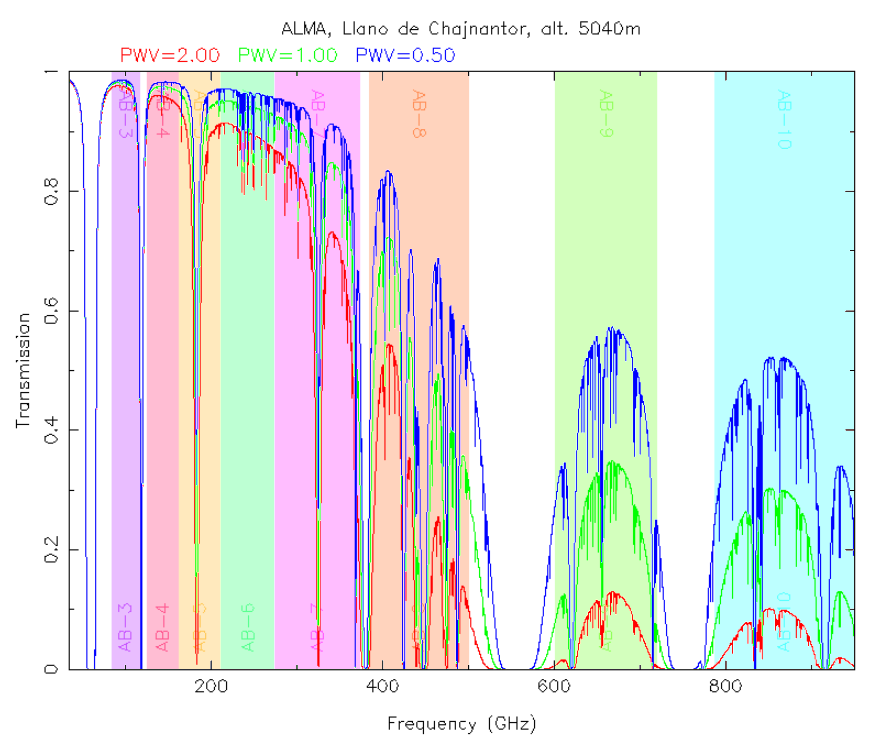}
    \caption{Atmospheric transmission at the ALMA site on Llano de Chajnantor, at different frequencies, for three values of precipitable water vapour (PWV): 2.0mm (red), 1.0mm (green) and 0.5mm (blue). The vertical shaded areas show the frequency coverage of the different ALMA Bands, from AB-3 to AB-10. Prediction obtained with the on-line calculator at \url{https://almascience.nrao.edu/about-alma/atmosphere-model}.}\label{fig:alma_transmission}
\end{figure}

Fig. \ref{fig:alma_transmission} shows the atmospheric transmission at the ALMA site on Llano de Chajnantor. The transmission curve is obtained from the atmospheric radiative transfer model for ALMA by \citet{pardo2019}, and covers the { 85} to 950 GHz frequency interval, equivalent to the 300$\mu$m to 3.6mm wavelength range. The figure shows the zenith atmospheric transmission for a precipitable water vapour (PWV) of 2.0mm, 1.0mm and 0.5mm. The PWV indicates the depth of water in the atmospheric column, if all the water in that column were measured as rainfall in mm. The PWV at the ALMA site is typically below 2.0mm for 65$\%$ of the year, below 1.0mm for 50$\%$ of the time and goes below 0.5mm for 25$\%$ of the time. Bands at higher frequencies (Band 9 and 10) are more affected by atmospheric transmission: considering a PWV of 1.0mm, the atmospheric transmission is $\sim$30$\%$ at 850 GHz (Band 10). Going toward shorter frequencies, the transmission goes up to $\sim$50$\%$ at 550GHz (Band 8), and $\sim$90$\%$ below 300GHz ({ Bands} 3-6). 

\begin{figure}
    \centering
    \includegraphics[width=0.90\textwidth]{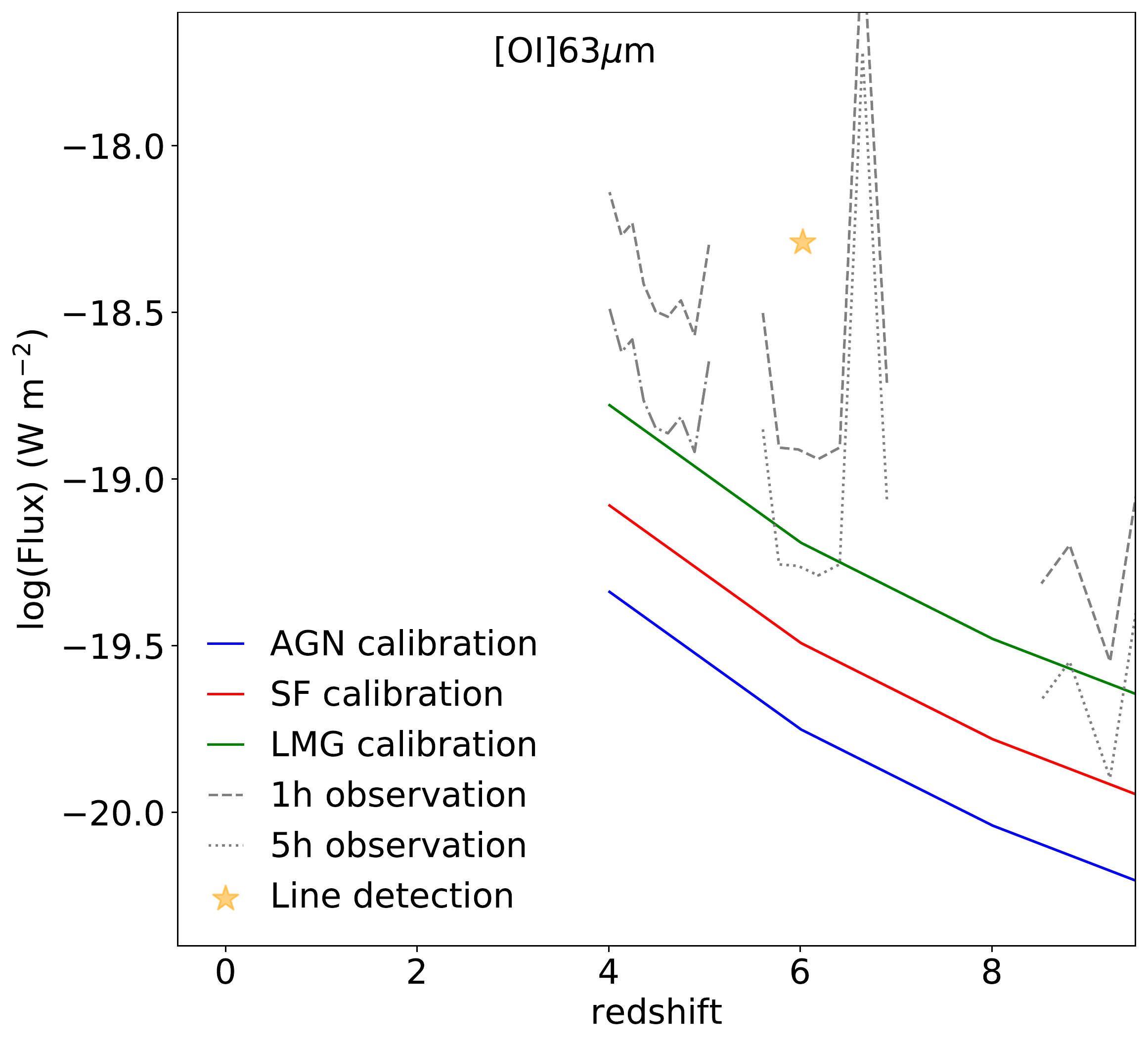}
    \includegraphics[width=0.90\textwidth]{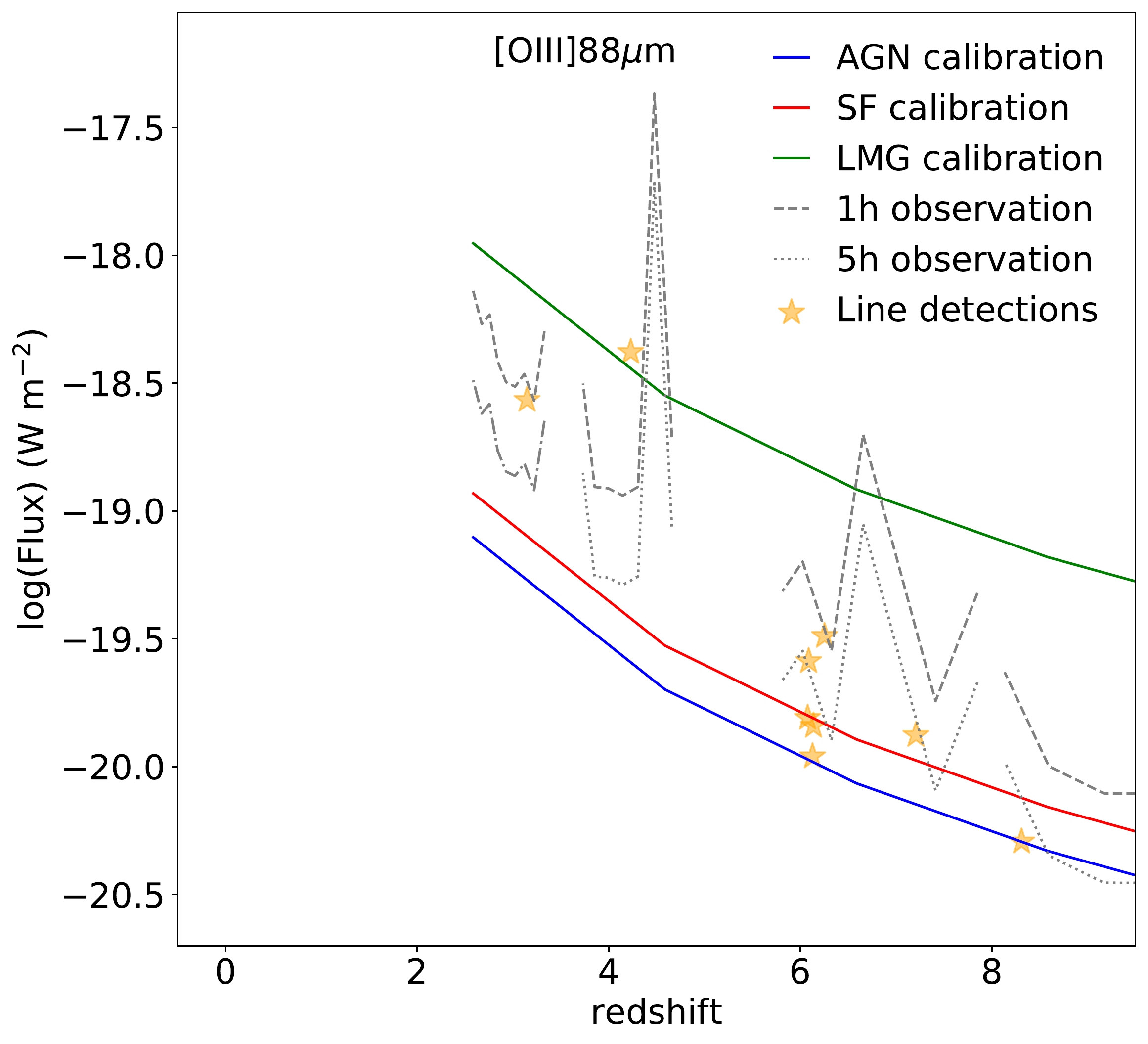}
    \caption{Predicted fluxes as a function of redshift for the [OI]63$\mu$m line (top) and the [OIII]88$\mu$m line (bottom), compared to the ALMA sensitivity for a 1 hr. (grey dashed line) and for a 5 hrs. observation (grey dotted line) up to redshift z$\sim$9. The blue solid line shows the predicted flux using the calibration for local AGN, the red line with the calibration for local SFG, and the green solid line with the one for local LMG. The various atmospheric absorption peaks show redshift intervals that cannot be observed. The orange stars show detections for each line. }
    \label{fig:alma_sensitivity_o1_o3}
\end{figure}

\begin{figure}
    \centering
    \includegraphics[width=0.90\textwidth]{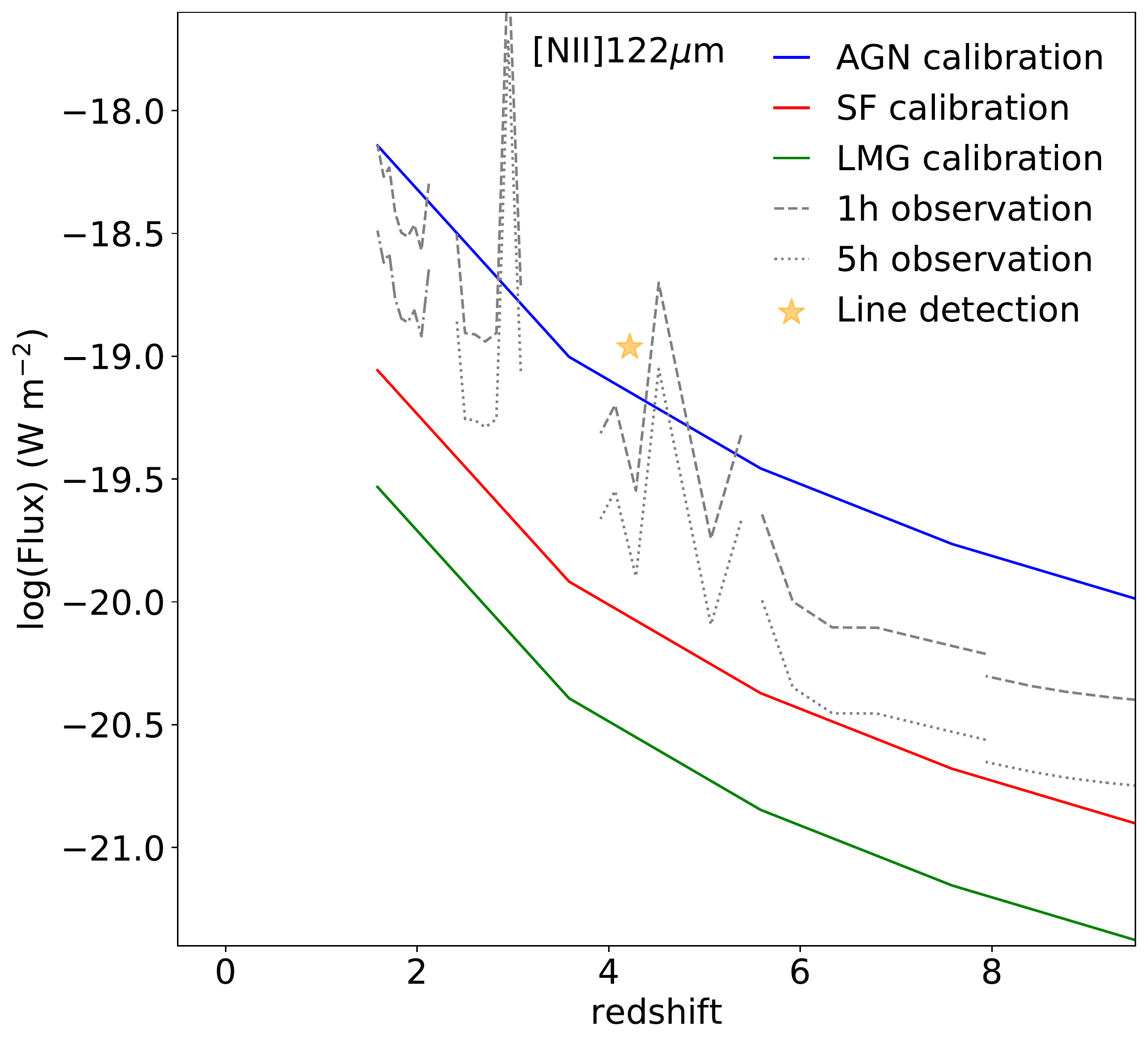}
    \includegraphics[width=0.90\textwidth]{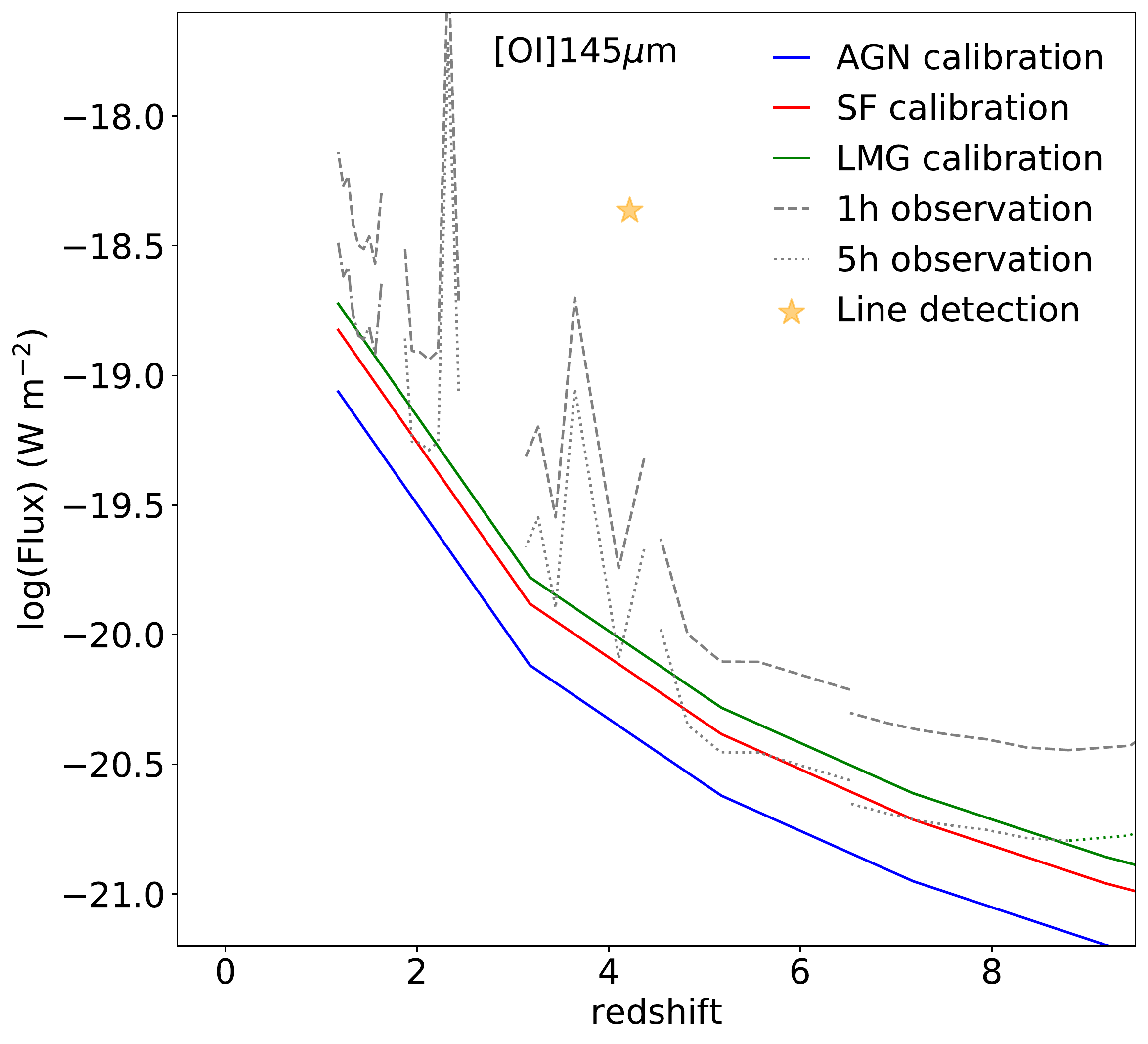}
    \caption{Predicted fluxes as a function of redshift for the [NII]122$\mu$m line (top) and the [OI]145$\mu$m line (bottom), compared to the  ALMA sensitivity for a 1 hr. (grey dashed line) and for a 5 hrs. observation (grey dotted line) up to redshift z$\sim$9. The blue solid line shows the predicted flux considering the calibration for local AGN, the red line shows the calibration for local SFG, and the green solid line shows the calibration for local LMG. The orange stars show detections for each line.}
    \label{fig:alma_sensitivity_n2_122_o1_145}
\end{figure}

\begin{figure}
    \centering
    \includegraphics[width=0.90\textwidth]{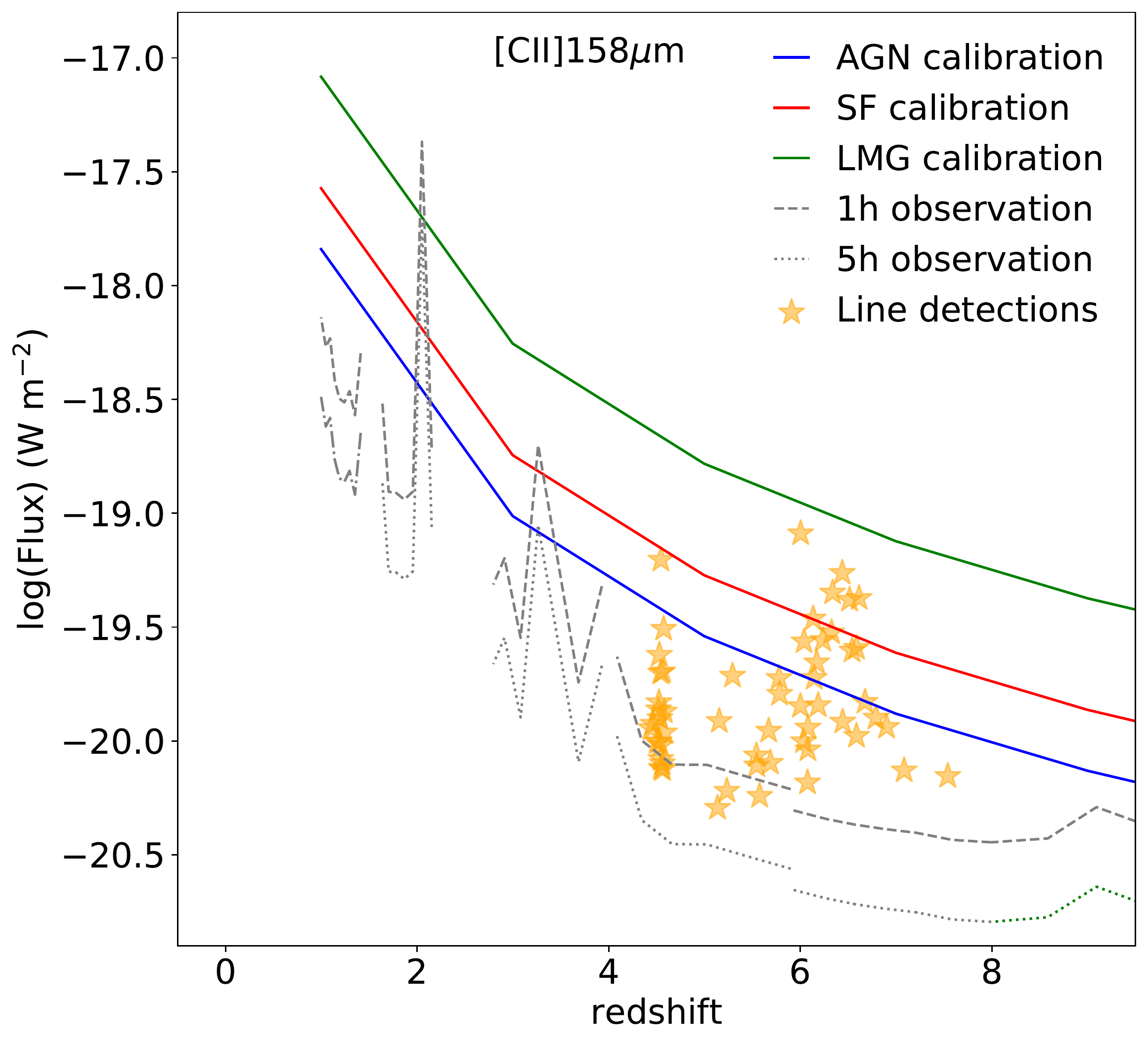}
    \includegraphics[width=0.90\textwidth]{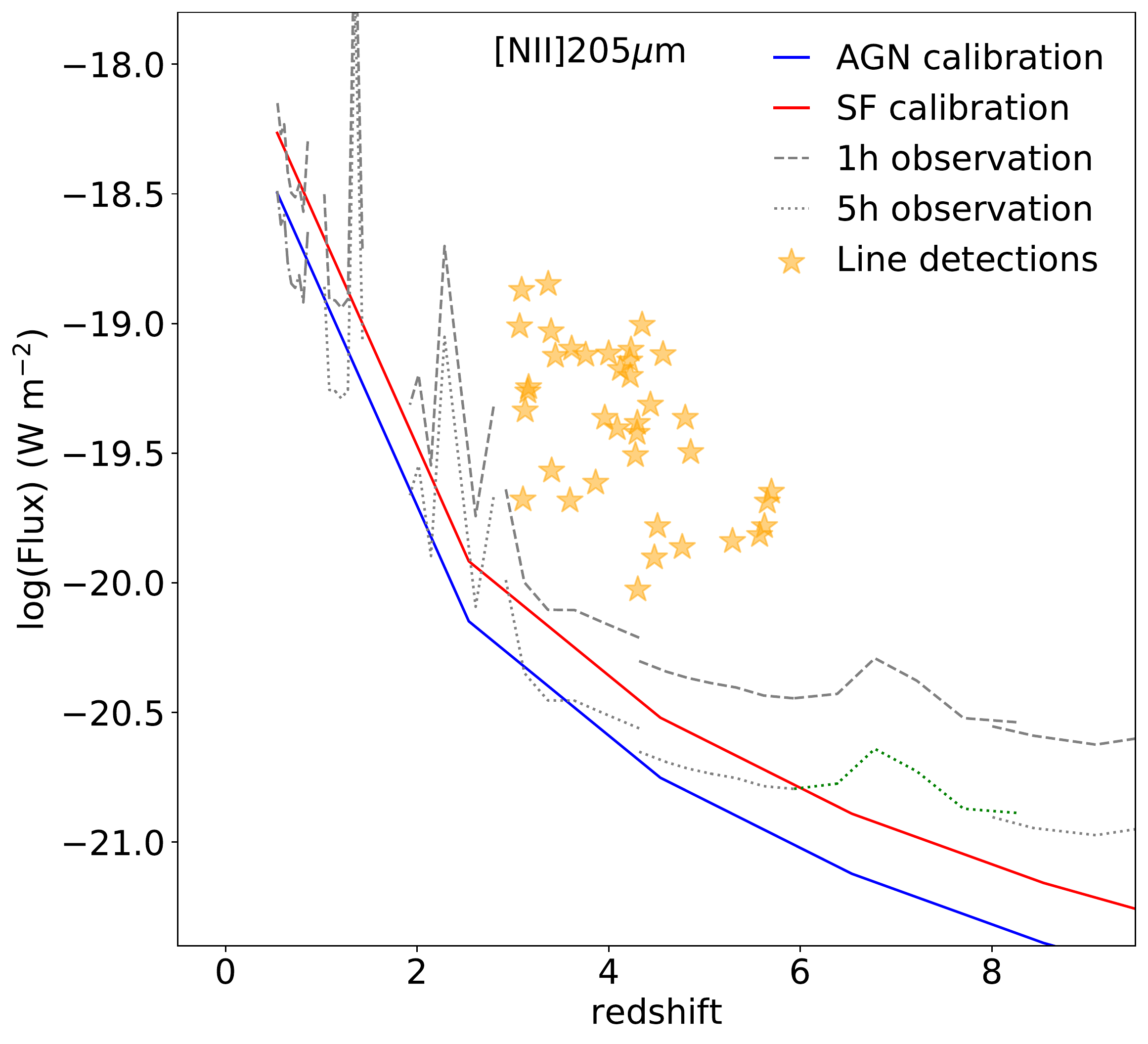}
    
    \caption{Predicted fluxes as a function of redshift for the [CII]158$\mu$m line (top) and the [NII]205$\mu$m line (bottom), compared to the ALMA sensitivity for a 1 hr. observation (grey dashed line) and for a 5 hrs. observation (grey dotted line) up to redshift z$\sim$9. The blue solid line shows the predicted flux considering the calibration for local AGN, the red line shows the calibration for local LMG. The orange stars show detections for each line.}
    \label{fig:alma_sensitivity_c2_n2_205}
\end{figure}

When determining the integration time using the ALMA sensitivity calculator\footnote{\url{https://almascience.eso.org/proposing/sensitivity-calculator}}, the PWV is automatically selected by the on-line tool to the most representative value for each band of observation.

As can be seen from Fig. \ref{fig:alma_sensitivity_o1_o3}-\ref{fig:alma_sensitivity_c2_n2_205} it has to be noticed that the ALMA sensitivity as a function of redshift is not a simple monotonic function, but contains discontinuities due to atmospheric absorption at particular frequencies. These figures show the fluxes predicted for various lines as a function of redshift for a source of total IR luminosity of L$_{IR}$=10$^{12.5}$L$_{\odot}$, compared to the ALMA sensitivity in 1 hr. and 5 hrs. reaching a signal-to-noise { SNR}=5, assuming the line calibrations for SFG, LMG and AGN presented in \citet{mordini2021}. Where available, we have included in the figures the detections so far reported in high redshift galaxies (indicated as yellow stars: $\star$). For predicting the fluxes at each redshift, we have chosen the luminosity of L$_{IR}$=10$^{12.5}$L$_{\odot}$ because it corresponds to the luminosity of a { Main-Sequence (MS)} galaxy of mass of M$_{\star}$=10$^{10.7}$ M$_{\odot}$ at a redshift z=4 \citep{scoville2017}. This allows us to predict the observability of galaxies at the knee of the luminosity and mass function at redshift z$\gtrsim$3.

In Fig.\,\ref{fig:alma_sensitivity_o1_o3} we show the fluxes predicted for the [OI]63$\mu$m (top) and [OIII]88$\mu$m (bottom) lines as a function of redshift. The [OI]63$\mu$m line could in principle be observed by ALMA starting from the redshift of z$\sim$4 and up to a redshift of z$\sim$8. However, the line is observable only using ALMA bands 9 and 10, making it rather difficult to detect, due to the low atmospheric transmission. Considering the LMG calibration, the line would require an integration time of $\sim$5 hrs. to be detected, while for the SFG and AGN calibrations the observational time would be between 10 and 20 hrs. This value is unrealistic for a ground-based telescope, such as ALMA, and in principle could be achieved only by summing up the results of various observing runs of few hours each. As a matter of fact, the only detection of the [OI]63$\mu$m line available so far for a high-redshift galaxy is the one reported by \citet{rybak2020}, { shown in the figure}, for a strongly gravitationally lensed galaxy (magnification $\mu_{FIR}\simeq$9) obtained with the APEX 12m telescope \citep{gusten2006}, and the Swedish ESO PI (SEPIA) Band 9 receiver \citep{belitsky2018}. 

The [OIII]88$\mu$m line, on the other hand, can be observed from redshift z$\sim$3. For this line, galaxies that present physical characteristics similar to local LMG can be easily detected with observations shorter than 1 hr., while for SFG and AGN, observational times longer than 5 hrs. are required up to redshift z$\sim$8. In the figure, we report the [OIII]88$\mu$m detections by \citet{vishwas2018,walter2018,debreuck2019,tamura2019,harikane2020}. We note that, while all reported detections have { signal to noise ratio of SNR}=5 or higher, the integration times can reach values of the order of 7.5 hrs. Moreover, we do not correct the line luminosities for the magnification effect, when present.

In Fig.\,\ref{fig:alma_sensitivity_n2_122_o1_145} we show the fluxes predicted for the [NII]122$\mu$m (top) and [OI]145$\mu$m (bottom) lines as a function of redshift.
In AGN, the [NII] line can be detected with an integration time of less than 1 hr., from a redshift of z$\sim$1.8. For SFG, the line can be detected with observations of the order of 5 hrs. only for redshifts above z$\sim$6, while at lower redshifts, or for LMG, the observational times needed to detect this line are { longer} than 5 hrs. The [OI]145$\mu$m line  can be observed with integration times of the order of 5hr. above redshift z$\sim$4 in SFG and LMG, while in AGN or for lower redshifts, longer integration times are needed. In the figure we also report a detection for [NII]122$\mu$m and [OI]145$\mu$m \citep{debreuck2019}. 

It is important to note, when considering nitrogen and oxygen lines, that the proposed predictions are derived for local galaxies, where the nitrogen to oxygen ratio shows typically values of about log(N/O)$\sim -0.6$ \citep{pilyugin2014}. When considering high redshift objects, however, this ratio may vary \citep[e.g.][]{amorin2010}, since it depends, among other parameters, on the stellar initial mass function, the star formation efficiency, and the star formation history due to the secondary origin of nitrogen in the CNO cycle in intermediate-mass stars. For this reason, while the detections of high redshift objects are in agreement with our calibration, different N/O ratios from the solar value log(N/O)$\sim -0.6$ would introduce an additional scatter in the calibration of the nitrogen lines for galaxies at high redshift.

Finally, Fig.\,\ref{fig:alma_sensitivity_c2_n2_205} shows the predicted fluxes for the [CII]158 $\mu$m (top) and the [NII]205$\mu$m (bottom) lines. The [CII] line is by far the brightest far-IR fine-structure line in galaxies and given its longer wavelength it can be detected in the redshift ranges of 0.9$\lesssim$z$\lesssim$2 and 3$\lesssim$z$\lesssim$9, requiring integration times shorter than 1 hr. independently of the galaxy type. The [NII]205$\mu$m line can be detected from redshift z$\sim$0.5, however it requires at least 1 hr. of observation up to z$\sim$3, and a longer integration at higher redshifts for SFG. For AGN the line appears weaker, and requires integration times longer than 5 hrs. to be detected. We include the detections for the [CII]158$\mu$m line reported by \citet{faisst2020,venemans2020} and for the [NII]205$\mu$m line by \citet{cunningham2020}, considering only objects with { signal to noise ratio of SNR}=5 or higher.

{ We note here that,} while in our calculations we have considered a source with a total IR luminosity of L$_{IR}$=10$^{12.5}$L$_{\odot}$, brighter galaxies, or sources whose apparent flux is enhanced by gravitational lensing, can be detected in shorter integration times.

\section{Measuring the star formation rate and the black hole accretion rate}
\label{sec:sfr_bhar}

\begin{figure*}
    \centering
    \includegraphics[width=0.456\textwidth]{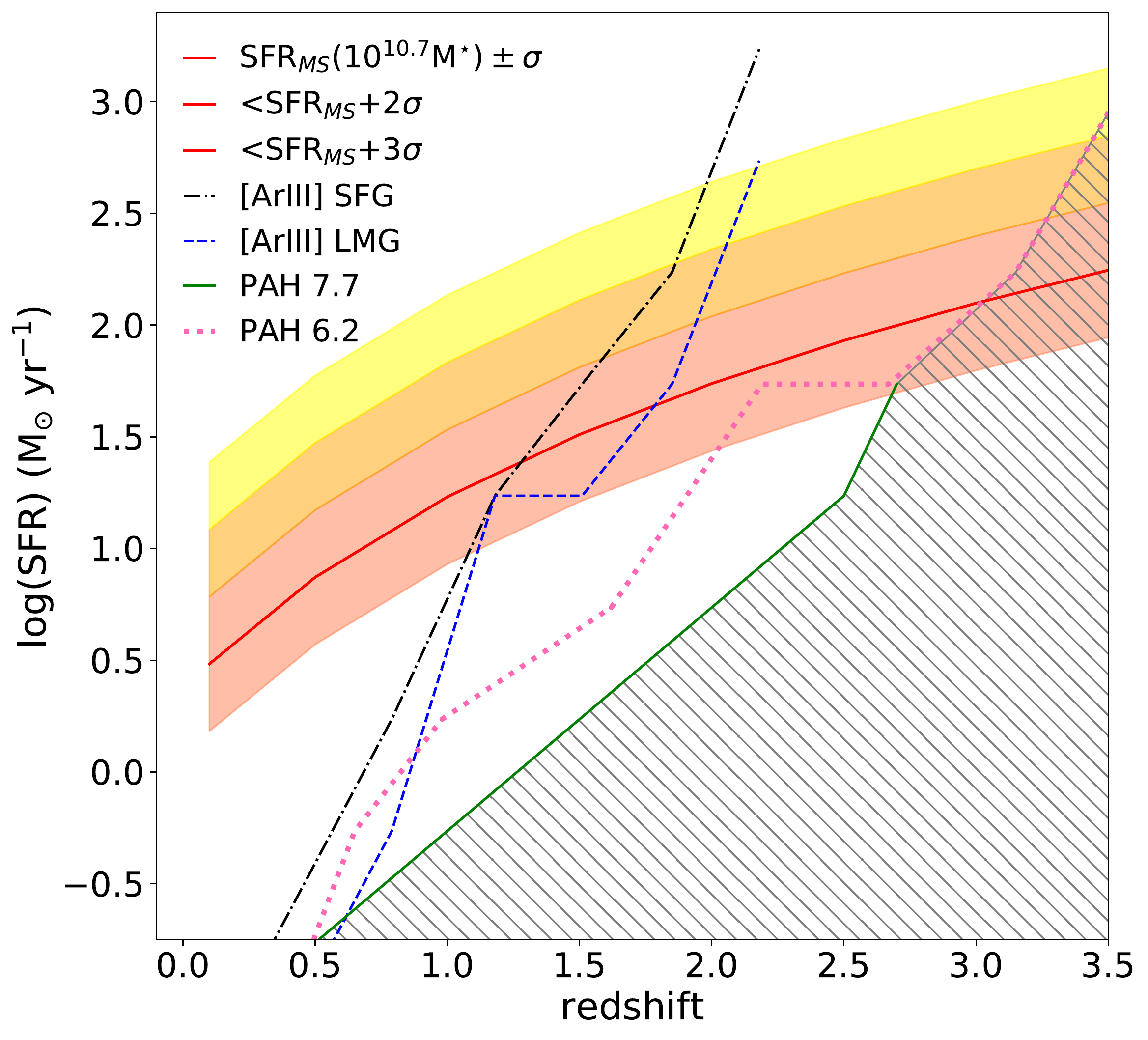}
    \includegraphics[width=0.44\textwidth]{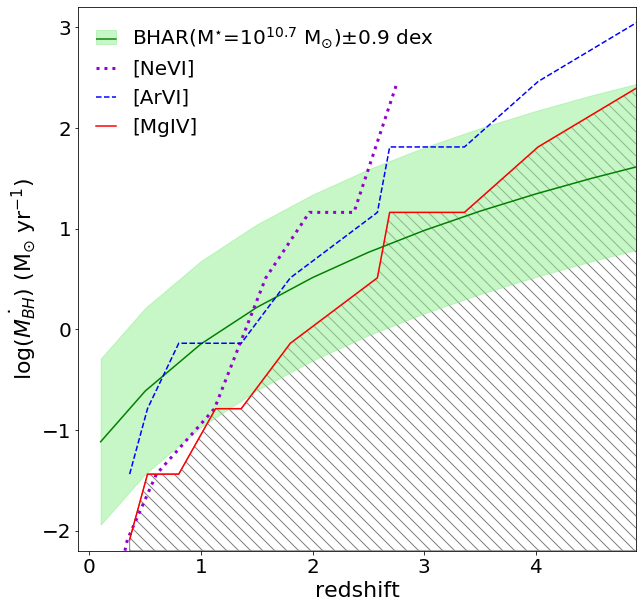}
    \caption{ ({\bf a: left}) Star formation rate (SFR, in M$_{\odot}$\,yr$^{-1}$) as a function of redshift for a 10$^{10.7}$ M$_{\odot}$ galaxy in the Main Sequence \citep[MS,][]{scoville2017} as a red solid line. The red-shaded area shows the $\sigma  =$\,0.35\,dex intrinsic scatter around the MS \citep{schreiber2015}, while the dark- and light-orange shaded areas indicate the +2$\sigma$ and +3$\sigma$ lines above the MS, respectively. The observability limits for the lines are shown as: black dash-dotted line: [ArIII]8.99$\mu$m for a SFG with t$_{INT}$ =1hr; blue dashed line: same line for a LMG (in both cases the highest line ratio present in Table \ref{tab:line_ratios} was adopted);  pink dotted line: the PAH feature at 6.2$\mu$m; green solid line: the PAH feature at 7.7$\mu$m. In these latter cases, we derived the limit from the correlations in \citet{mordini2021}. ({\bf b: right}) The green solid line indicates the instantaneous BH accretion rate (BHAR, in M$_{\odot}$\,yr$^{-1}$) as a function of redshift expected for a MS galaxy with a mass of 10$^{10.7}$ M$_{\odot}$ during its active BH accretion phase at each epoch, using the SFR-BHAR relation of \citet{diamondstanic2012}. The green shaded area shows the associated dispersion. The observability limits for the lines are shown as: purple dotted line: the [NeVI]7.65$\mu$m line; blue dashed line: the [ArVI]4.53$\mu$m; red solid line: the [MgIV]4.49$\mu$m line. For all three lines, we considered a 1hr. pointed observation, and the highest line ratio as reported in Table \ref{tab:line_ratios}. In both panels, the grey hatched areas indicate the parameter space that requires observations longer than 1hr.}
    \label{fig:sfr_bhar_miri}
\end{figure*}

We derive here the prescriptions to measure the SFR and the BHAR with the JWST-MIRI spectrometer at the Cosmic Noon (1$\lesssim $z$\lesssim$3), and the SFR with ALMA at higher redshift (z$\gtrsim$3).

For the JWST predictions, we follow the same method that has been prepared for the scientific assessment of the galaxy evolution observational program \citep{spinoglio2021} for the SPICA mission \citep{roelfsema2018}. In this study, the predictions of measuring the SFR and the BHAR with the 2.5m SPICA telescope have been carried out using the [NeII]12.8$\mu$m and [NeIII]15.6$\mu$m lines for the SFR and [OIV]26$\mu$m for the BHAR \citep[see fig.\,9 of][]{spinoglio2021}, while for JWST-MIRI we use shorter wavelength lines.

Fig.\ref{fig:sfr_bhar_miri}(a) shows the SFR of a galaxy in the Main Sequence \citep[MS;][]{rodighiero2010,elbaz2011}, with values of luminosities as a function of redshift taken from \citet{scoville2007}, assuming a galaxy with a stellar mass of M$_{\star}$=10$^{10.7}$ M$_{\odot}$ \citep{muzzin2013,adams2021} and an associated dispersion around the MS of 0.35 dex \citep[red shaded area;][]{schreiber2015}. The relation between luminosity and SFR has been taken from \citet{kennicutt2012}. The various colored lines indicate the JWST-MIRI sensitivity limit for a 1 hr. pointed observation, considering the PAH features at 6.2 and 7.7$\mu$m for SFG and the [ArIII]8.99$\mu$m line for both SFG and LMG, as discussed in Section \ref{sec:results_jwst}. For the PAH features, we considered the calibrations of  \citet{mordini2021}, while for the [ArIII] line calibration we adopted the higher line ratio to the [NeII] line, { whose calibration is given in \citet{mordini2021}}, reported in Table \ref{tab:line_ratios}, and thus the physical conditions yielding the brightest lines. The PAH features at 6.2$\mu$m and 7.7$\mu$m are the best tracers for the SFR, covering the population below the MS up to redshift z$\sim$3, with a minimum luminosity of L$_{IR}\sim 10^{9.5}\, \rm{L_\odot}$ at z$\sim$1 and L$_{IR}\sim 10^{12.5}\, \rm{L_\odot}$ at z$\sim$3. The [ArIII]8.99$\mu$m line, on the other hand, can be detected in MS galaxies up to redshift z$\sim$1.8 (limiting luminosities between L$_{IR}\sim 10^{11}\, \rm{L_\odot}$ at z$\sim$1 and L$_{IR}\sim 10^{13}\, \rm{L_\odot}$ at z$\sim$2.2). This poses a limit to the study of LMG at higher redshifts, since the reduced formation efficiency and the increased stellar radiation hardness reduces the strength of the PAH features in these { galaxies}, which, { in the local universe}, are only detected at metallicities above 1/8-1/10 Z$_{\odot}$ \citep{engelbracht2008,cormier2015,galliano2021}. 

The results presented for the ALMA telescope suggest that the [CII]158$\mu$m line, a proxy for measuring the SFR, can be easily observed in objects at redshift z$\gtrsim$3, thus allowing to trace the evolution of SFR throughout cosmic time. At the present time, this line has been observed by ALMA in the z$\sim$4-7 redshift interval \citep[see e.g. ][]{hashimoto2019,faisst2020}. An alternative can be offered by the sum of the [OIII]88$\mu$m and [OI]145$\mu$m lines, in analogy to the [OI]63$\mu$m plus [OIII]88$\mu$m tracer proposed in \citet{mordini2021}. { Because of its longer wavelength, the [OI]145$\mu$m line is not affected by the strong atmospheric absorption that makes very difficult the observations of the [OI]63$\mu$m line, as shown in Figs. \ref{fig:alma_sensitivity_o1_o3} and \ref{fig:alma_sensitivity_n2_122_o1_145}.}
The [OI]145$\mu$m can be detected with ALMA above redshift z$\sim$1.5 with observations of $\sim$5\,hr for SFG and LMG-like sources, and, unlike the [OI]63$\mu$m line, is not affected by self-absorption \citep[see, e.g.,][]{liseau2006}. 

We report here the correlations between the SFR and the PAH 6.2$\mu$m feature luminosity (equation \ref{eq:pah_sfr}), that can be observed by the JWST-MIRI, and between the SFR and the [CII]158$\mu$m line luminosity (equation \ref{eq:c2_sfr}), currently observed by ALMA, derived by \citet{mordini2021}:

\begin{multline}\label{eq:pah_sfr}
    \log\left(\frac{SFR}{\rm M_{\odot}\,yr^{-1}}\right)=(0.37 \pm 0.04)\\+(0.76 \pm 0.03)\log\left(\frac{L_{\rm PAH6.2}}{\rm 10^{41}\,erg\,s^{-1}}\right)
\end{multline}

\begin{multline}\label{eq:c2_sfr}
\log\left ( \frac{SFR}{\rm M_{\odot}\,yr^{-1}}\right )=(0.62 \pm 0.02)\\ +(0.89 \pm 0.02) \left( \log\frac{L_{\rm [CII]}}{\rm 10^{41}\,erg\,s^{-1}}\right )
\end{multline}

with the line luminosities expressed in units of 10$^{41}$\,erg\,s$^{-1}$, and the SFR in M$_{\odot}$\,yr$^{-1}$. We note that in local SFG, the bulk of the [CII] emission arises from neutral gas \citep{croxall2017}, with only a minor contribution from ionized gas, as discussed also in Section \ref{sec:other_components}. In LMG, however, PDR regions are less frequent, and the [CII] emission arises mainly from ionized regions \citep[see e.g. ][]{cormier2019}. The two components, however, compensate each other, thus making this line an ideal tracer for star formation activity in galaxies with different ISM conditions.

In analogy with the [NeII]12.8$\mu$m plus [NeIII]15.6$\mu$m SFR tracer \citep[see Section 3.2.4 in][ for details]{mordini2021}, we derive the correlation between the SFR and the sum of the [ArII]6.98$\mu$m and [ArIII]8.99$\mu$m lines, considering the ([ArII]+[ArIII])/([NeII]+[NeIII]) ratio (reported in Table \ref{tab:line_ratios}), as computed from the photoionization models. This leads to a SFR to ([ArII]+[ArIII]) lines luminosity correlation for SFG and LMG reported in equation \ref{eq:ar2_ar3_sfr_sfg} and \ref{eq:ar2_ar3_sfr_lmg}, respectively:

\begin{equation}\label{eq:ar2_ar3_sfr_sfg}
 \log\left(\frac{SFR}{\rm M_{\odot}\,yr^{-1}}\right)= 1.2+0.96 \log\left( \frac{L_{\rm [ArII]+[ArIII]}}{\rm 10^{41}\,erg\,s^{-1}}\right)
\end{equation}

\begin{equation}\label{eq:ar2_ar3_sfr_lmg}
 \log\left(\frac{SFR}{\rm M_{\odot}\,yr^{-1}}\right)= 1.57+0.96 \log\left( \frac{L_{\rm [ArII]+[ArIII]}}{\rm 10^{41}\,erg\,s^{-1}}\right)
\end{equation}

Here we use the composite line ratio of ([ArII]+[ArIII])/ ([NeII]+[NeIII]) instead of ratios of single lines, to be able to cancel out the dependence on the metallicity, whose lower value compared to solar would increase the higher ionization line compared to the lower ionization one. The SFR tracer derived in this way is independent of metallicity.  

\begin{figure}
    \centering
    \includegraphics[width=0.85\columnwidth]{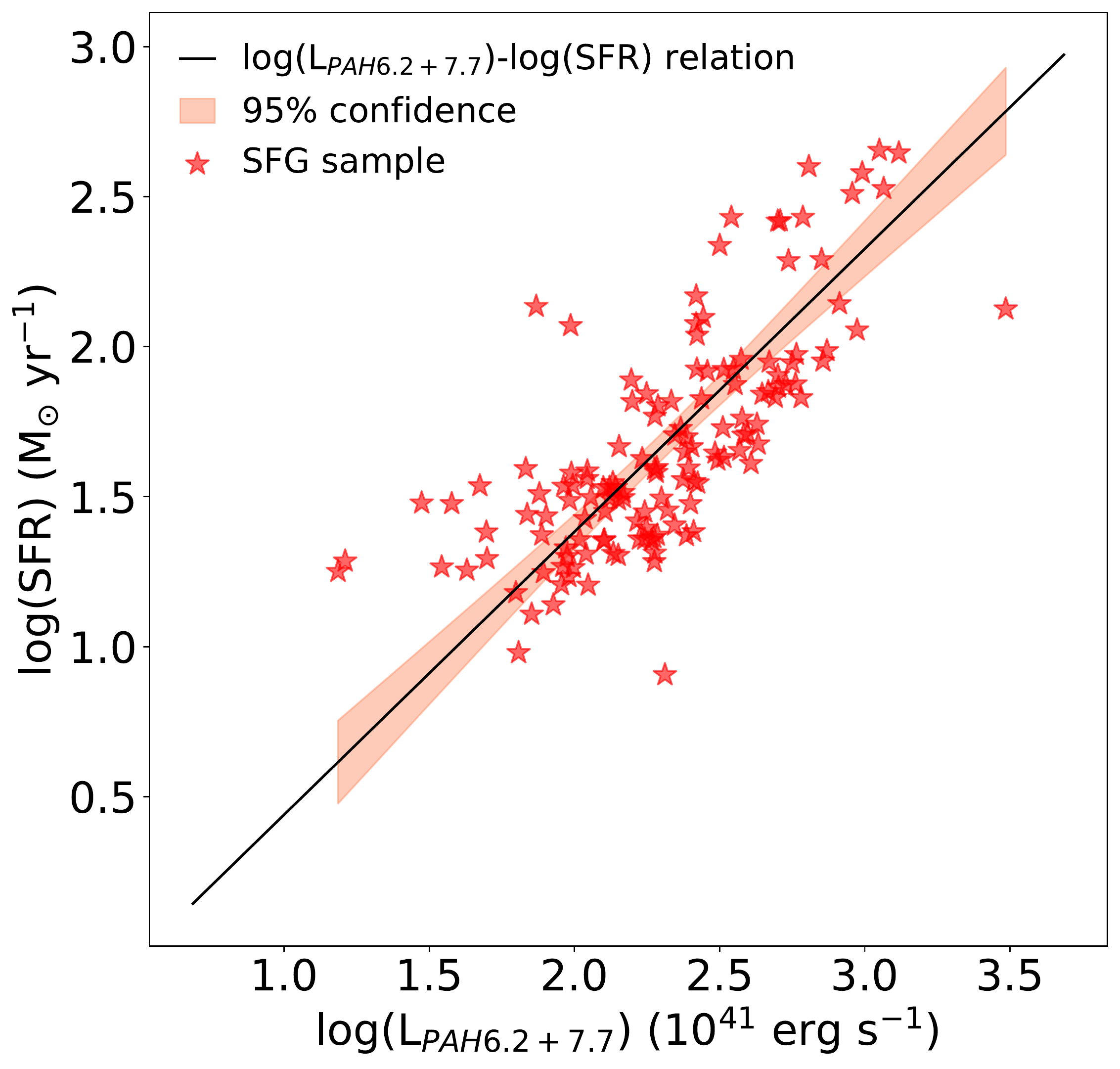}
    \caption{Correlation between the summed PAH features luminosities at 6.2 and 7.7$\mu$m, in units of $10^{41}\, \rm{erg\,s^{-1}}$, and the SFR derived from the total IR luminosity (black solid line) for a sample of SFGs (red stars). See section 2 in \citet{mordini2021} for details on the sample.}
    \label{fig:sfr_pah}
\end{figure}

In a similar way, while the single PAH feature might adequately trace SFR, a combination of two o more features can better keep into account the different conditions of the ISM of star forming regions. We propose here a combination of the 6.2+7.7$\mu$m PAH features to measure the SFR. The calibration, derived using 139 SFG galaxies, is shown in Fig. \ref{fig:sfr_pah} and is described by the following equation:

\begin{multline}
\log\left(\frac{SFR}{\rm M_{\odot}\,yr^{-1}}\right)=(-0.50 \pm 0.15)\\+(0.94 \pm 0.06)\log\left(\frac{L_{\rm PAH6.2+7.7}}{\rm 10^{41}\,erg\,s^{-1}}\right)        
\end{multline}

The combination of the two PAH features can be used to measure SFR up to redshift z$\sim$2.7.

\begin{figure}[!hb]
    \centering
    \includegraphics[width=0.85\columnwidth]{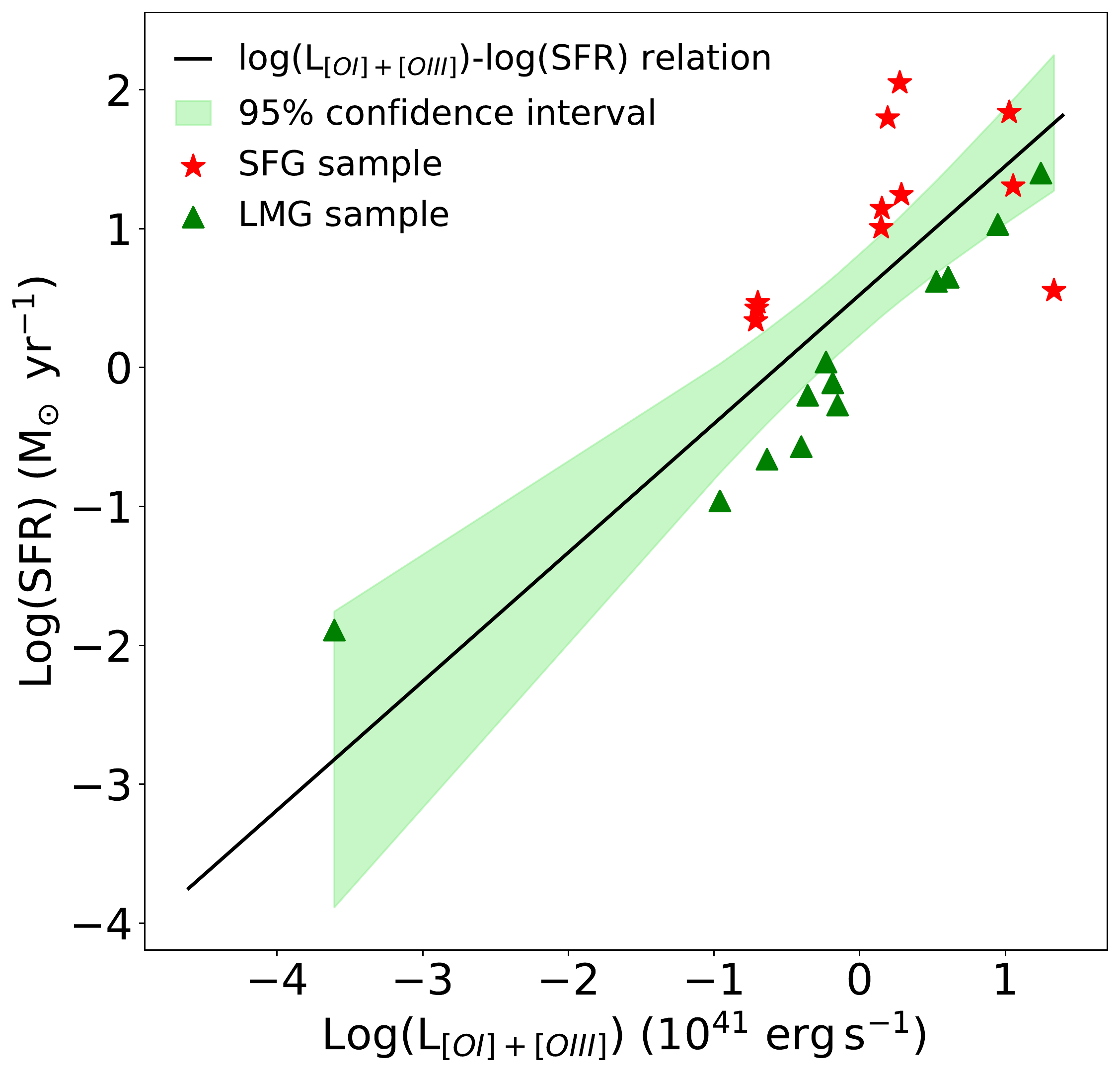}
    \caption{Correlation between the [OIII]88$\mu$m and [OI]145$\mu$m summed emission line luminosities, in units of $10^{41}\, \rm{erg\,s^{-1}}$, and the SFR derived from the total IR luminosity (black solid line) for a composite sample of SFGs (red stars) and from the H$\alpha$ luminosity (corrected for the IR luminosity) for LMG (green triangles). See section 3 in \citet{mordini2021} for details on the determination of the SFR.}
    \label{fig:o1_o3_sfr}
\end{figure}

Following \citet{mordini2021}, who proposed the use of the sum of the [OI]63$\mu$m plus [OIII]88$\mu$m lines luminosity as a SFR tracer, we computed, for local galaxies, a correlation between the SFR and the sum of the [OIII]88$\mu$m plus [OI]145$\mu$m lines luminosity. 
Fig.\,\ref{fig:o1_o3_sfr} shows the correlation described by the equation:

\begin{multline}
 \log\left(\frac{SFR}{\rm M_{\odot}\,yr^{-1}}\right)=(0.52 \pm 0.14)\\+(0.93 \pm 0.14) \log\left( \frac{L_{\rm [OI]+[OIII]}}{\rm 10^{41}\,erg\,s^{-1}}\right)
\end{multline}

The correlation was obtained using 23 local galaxies, of which 11 SFG \citep[from][]{negishi2001} and 12 LMG \citep[from][]{cormier2015}, and has a Pearson $r$ coefficient of 0.77. The relatively small number of objects used to derive this correlation is due to the few detections of the [OI]145$\mu$m line in local galaxies, which broadens the confidence interval of the relation, reported in Fig.\,\ref{fig:o1_o3_sfr} as a shaded area.

Fig.\ref{fig:sfr_bhar_miri}(b) shows the instantaneous BHAR, that can be measured by JWST-MIRI, as a function of redshift for an AGN in a MS star-forming galaxy with M$^{*}$=10$^{10.7}$ M$_{\odot}$, using the SFR-BHAR calibration derived by \citet{diamondstanic2012} for a sample of nearby Seyfert galaxies. The green shaded area corresponds to the dispersion of the SFR-BHAR relation, i.e. $\sim$ 0.9 dex. This estimate, however, considers only the portion of galaxies in the MS undergoing a BH accreting phase at each epoch, with the instantaneous BHAR that has not been averaged over the duty cycle of the active nucleus. The dashed lines represent the three proposed tracers for BHAR in Section \ref{sec:results_jwst}, considering a 1 hr. pointed observation: the [MgIV]4.49$\mu$m line appears to be the best tracer for BHAR, reaching galaxies below the main sequence up to redshift z$\sim$2.5 (observable luminosities between L$_{IR}\sim 10^{10.5}\, \rm{L_\odot}$ at z$\sim$1 and L$_{IR}\sim 10^{12.5}\, \rm{L_\odot}$ at z$\sim$4). The possibility of detecting the [ArVI] and [NeVI] lines only in bright galaxies is due to the higher ionization potential needed to excite these lines (respectively of 80\,eV and 126\,eV, see Table \ref{tab:line_properties}). 


It is important to consider that the lines analyzed in this work have not been extensively observed yet, and these predictions are based on CLOUDY photo-ionization models. Considering the line ratios reported in Table \ref{tab:line_ratios} and the BHAR-[NeV]14.3$\mu$m correlation derived in \citet{mordini2021} (see Section 3.3 and Appendix C there for details on the derivation of the correlation), we obtain a correlation between the BHAR and the [MgIV]4.49$\mu$m line (reported in equation \ref{eq:mg4_bhar}), and between the BHAR and the [ArVI]4.53$\mu$m line (reported in equation \ref{eq:ar6_bhar}).  

\begin{equation}\label{eq:mg4_bhar}
    \log\left(\frac{\dot{M}_{BH}}{\rm M_{\odot}\,yr^{-1}}\right)=-0.85+1.04\log\left(\frac {L_{\rm [MgIV]4.49}}{\rm 10^{41}\,erg\,s^{-1}}\right)
\end{equation}

\begin{equation}\label{eq:ar6_bhar}
\log\left(\frac{\dot{M}_{BH}}{\rm M_{\odot}\,yr^{-1}}\right)=0.58+1.04\log\left(\frac {L_{\rm [ArVI]4.53}}{\rm 10^{41}\,erg\,s^{-1}}\right)
\end{equation}

The hatched areas in both panels of Fig.\ref{fig:sfr_bhar_miri} indicate the region where observations longer than 1 hr are required. A longer exposure time can significantly improve these limits, allowing the JWST-MIRI spectrometer to study both SFR and BHAR in MS galaxies up to redshift z$\sim$3. 

Among the far-IR lines covered by the ALMA telescope at redshifts higher than z$\sim$3, the only available tracers for BH activity are represented by the [OIII]52 and 88$\mu$m lines. Their ionization potential ($\sim$35\,eV, see Table \ref{tab:line_properties}) is sufficiently high to be excited by AGN. However, these lines are also associated to stellar and HII regions excitation, thus limiting their potential as BHA tracer. A { large} set of observed far-IR lines { and} specific photo-ionization models that could simulate the two emission components \citep[see, e.g.][{ for the case of NGC1068}]{spinoglio2005} are needed to disentangle the stellar emission from the total observed emission of these lines.

\subsection{Applications to composite objects}\label{sec:composite_objects}
The coexistence of AGN and star formation in galaxies is well known \citep[see, e.g.][for a review]{perez-torres2021} and therefore it should be expected that many galaxies are characterized by both components at work together. The high ionization lines considered in this work can only be produced by AGN activity, and can be used in composite objects to derive the BHAR. 
Mid-ionization lines, however, can be produced by both AGN and star formation activity, leading to possible overestimations of the SFR.

\begin{figure}
    \centering
    \includegraphics[width=0.95\columnwidth]{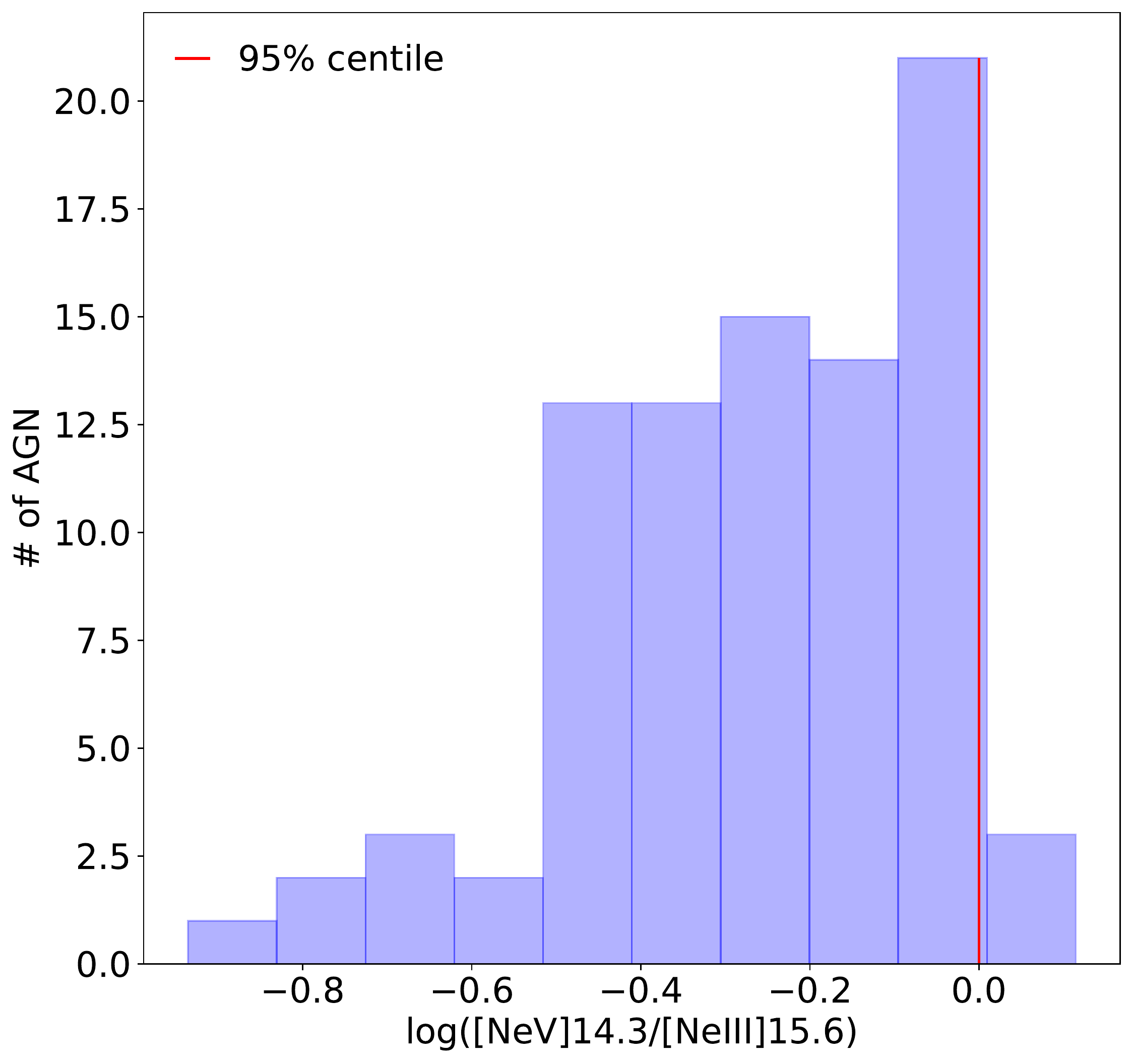}
    \caption{Histogram of the logarithm of the [NeV]14.3$\mu$m to [NeIII]15.5$\mu$m line for the \citet{tommasin2010} AGN sample. The vertical red line shows the 95\% percentile of the distribution.}
    \label{fig:agn_histogram}
\end{figure}

The combination of high- and mid-ionization lines, ideally of the same element, is necessary to estimate the relative contribution of each phenomenon to the total energy output in a galaxy. For instance, Fig.\ref{fig:agn_histogram} shows the [NeV]14.3$\mu$m to [NeIII]15.6$\mu$m ratio in the \citet{tommasin2010} AGN sample. Based on this ratio, we consider that a fully AGN-dominated source correponds to a line ratio of about [NeV]/ [NeIII]$\sim$0.9, which represents the 95$\%$ percentile of the line ratio distribution (vertical red line in Fig.\,\ref{fig:agn_histogram}). A lower line ratio indicates an increasing contribution from the star formation component, corresponds to 100$\%$ when [NeV] is not detected.

\citet{zhuang2019}, using mid-IR spectra and CLOUDY simulations, performed a similar analysis, deriving the average [NeII]12.8$\mu$m/[NeV]14.3$\mu$m and [NeIII] 15.5 $\mu$m/ [NeV]14.3 $\mu$m ratios, finding that AGN emit a relatively limited range of [NeII] and [NeIII] for a given [NeV] flux. This would allow one to predict the amount of AGN emission for these lines, and correct the observed flux in order to derive the component linked only to star formation activity.

While JWST will be able to detect both neon lines only up to redshift z$\sim$0.8, a similar analysis could be performed using the [ArVI]4.53 and [ArIII]8.99$\mu$m lines to reach higher redshift objects.

\subsection{Other components in galaxies: PDR and shocks  }\label{sec:other_components}

In this study, we have considered so far galaxies which are dominated by an AGN, by the star formation component (SFG) and also galaxies characterized by a low metallicity (LMG) using the spectroscopy that has been collected by the {\it Spitzer} and {\it Herschel} satellites. 
Moreover, we have shown in the previous section (\ref{sec:composite_objects}) how the AGN and star formation components can be disentangled and evaluated through the intermediate ionization lines. However, other emission components can be present in galaxies and in particular those originated in the so-called Photo Dissociation Regions \citep[PDR, ][]{tielens1985} and in shocks.

To disentangle the PDR emission in galaxies from the normal photoionized component, which we expect in the presence of HII regions emission, and therefore in the presence of star formation, we can use the ratio of the [CII]158$\mu$m line to one of the two [NII] lines, either at 122$\mu$m and 205$\mu$m. \citet{spinoglio2015} have shown that these line ratios, together with photoionization models simulating both a starburst (with the integration stopping at the temperature of 1000K) and a starburst plus a PDR model (with the integration going down to the temperature of 50K, including the neutral/low ionization gas contributing to the PDR), have a very different [CII]158$\mu$m/[NII]122$\mu$m and [CII]158$\mu$m/[NII]205$\mu$m line ratios. For the galaxies observed in \citet{spinoglio2015} most of the [CII] emission was originated in PDR, whilst only one fifth of the total emission was produced in pure HII regions. Thus, ALMA observations of at least one the [NII]122,205$\mu$m lines would be required to estimate the contribution of low-density ionized gas to the [CII]158$\mu$m emission and disentangling the PDR component. 

A similar approach, albeit more complex, can be used for disentangling the presence of shock excited emission. It is in fact well known, since the pioneering works of \citet{shull1979,hollenbach1989}, that dissociative shocks produce a plethora of intense atomic and ionic fine-structure lines. Notably the [OI]63$\mu$m line intensity is proportional to the particle flux (density $\times$ velocity) into a dissociative shock, because this line is the dominant coolant at temperature of T$\lesssim$5000K \citep{hollenbach1985}. Among the lines discussed in this study, besides [OI]63$\mu$m, also lines of [NeII]12.8$\mu$m, [SiII]34.8$\mu$m, 
and [OIII]52 and 88$\mu$m are present in fast dissociative shocks \citep{shull1979,hollenbach1989}.
To discriminate between shocks and PDR origin the ratio of [OI]63$\mu$m/
[CII]158$\mu$m can be used \citep{tielens1985}, although this line ratio should be used with caution since LMG galaxies also show high [OI]63$\mu$m/
[CII]158$\mu$m ratios \citep{cormier2019}.

In conclusion, if one wants to disentangle shock excitation from the intensities of observed fine-structure lines in galaxies, has first to choose an adequate shock excitation model able to predict their intensities and fit the results. These detailed computations are, however, beyond the scope of the present study, whose main aim is to give simple recipies to derive from the observed lines the SFR and the BHAR.

\section{Discussion}
\label{sec:discussion}

The low observational efficiency at performing wide area surveys (of the order of square degrees) with either of the two facilities, JWST and ALMA, imposes a significant limit to study galaxy evolution: while the tracers proposed in this work can be used to trace both star formation and BHA at the Cosmic Noon, a complete study of galaxy evolution also requires to account for the environment within which galaxies evolve, thus incorporating the large scale structure of the Universe, by mapping large cosmological volumes. Moreover, even using both facilities together, it is not possible to perform an accurate analysis of galaxy evolution covering both the Cosmic Noon and the higher redshifts. On the one hand, in fact, while the PAH features can be detected with JWST up to redshift z$\sim$2.5 in MS galaxies, and at higher redshifts the ALMA telescope can trace star formation activity with the [CII]158$\mu$m line, there is the problem of tracing SF in low metallicity environments, where the PAH features are weak or absent, or in high luminosity galaxies, where the [CII] deficit can lead to lower determinations of the SFR \citep[see e.g. ][]{ferrara2019}. On the other hand, ALMA lacks an unambiguous tracer for BHA activity, and the proposed tracers for JWST only reach redshift z$\sim$3 for MS galaxies.

A full characterization of the obscured phase of galaxy evolution at the Cosmic Noon will have to wait for a new IR space telescope, actively cooled down to a few degrees Kelvin and with state of the art detectors, able to perform wide area spectroscopic surveys covering most of the mid- to far-IR range (i.e. between 10$\mu$m and 300$\mu$m). Possible candidate missions include the very ambitious \textit{Origins Space Telescope} \citep[OST, ][]{meixner2019} and the smaller size \textit{Galaxy Evolution Probe} telescope \citep[GEP, ][]{glenn2021}, which has the main goal to map the history of galaxy growth through star formation and accretion by supermassive black holes and to characterize the relationship between these two processes. The pioneering study of the scientific potential of an IR space telescope for galaxy evolution studies throughout cosmic time has been provided by \citet{spinoglio2017, spinoglio2021}, based on the SPICA mission project \citep{swinyard2009, roelfsema2018}.

\section{Conclusions}
\label{sec:conclusions}

In this work we present spectroscopic tracers for star formation and BHA activity that can be used by the JWST and the ALMA telescopes to characterize galaxy evolution up to and beyond the Cosmic Noon, respectively. In particular, our findings can be summarized as follow:

\begin{itemize}
\item The JWST-MIRI instrument in spectroscopic mode will be able to detect mid-IR spectra of faint fine-structure lines and PAH features in galaxies at redshift z$\gtrsim$1 and discriminate between AGN-dominated and starburst-dominated galaxies and measure their metallicity.

\item The presence of an AGN can be detected using high ionization lines such as the [MgIV]4.49$\mu$m, the [ArVI]4.53$\mu$m or the [NeVI]7.65$\mu$m lines: the [MgIV] and [ArVI] lines can be observed up to redshift z$\sim$3, while the [NeVI] line can be detected with JWST-MIRI for a main-sequence galaxy at redshift z$\sim$1.5 in a 1 hr. observation. 

\item For the star formation processes using JWST, the PAH features can be easily detected up to redshift z$\sim$2.7, while the [ArII]6.98$\mu$m and [ArIII]8.99$\mu$m lines can be detected up to redshift z$\sim$2.5-3. The detection of the [SIV]10.5$\mu$m line can discriminate between solar and sub-solar metallicity: in LMG, the [SIV] line can be easily detected up to redshift z$\sim$1.5, while in solar metallicity SFG, this line is very weak and cannot be detected above redshift z$\sim$0.8.

\item Using the ALMA observatory, the far-IR oxygen lines of  [OIII]88$\mu$m and [OI]145$\mu$m can be detected in low-metallicity galaxies (LMG) with luminosities of about $10^{12.5}\, \rm{L_\odot}$ with integration times of 1 to 5 hrs. In AGN or normal SFG with similar luminosities, the [OIII]88$\mu$m line requires t$_{INT}$ $\gtrsim$ 5 hrs. Due to the high observing frequency, where the sky transparency is very low, the [OI]63$\mu$m can hardly be detected in t$_{INT}$ $\gtrsim$ 5 hrs. and only if the galaxy is a LMG. On the contrary the [NII]122$\mu$m line is preferentially detected in AGN and normal star forming galaxies (SFG). The [NII]205$\mu$m can hardly be detected with t$_{INT}$ $\gtrsim$ 5 hrs in AGN and SFG. 

\item The [CII]158$\mu$m line can be easily detected in any main-sequence galaxy at redshift of 0.9$\lesssim$z$\lesssim$2 and 3$\lesssim$z$\lesssim$9. It remains the best tracer for to measure the SFR, but it can also be easily detected in AGN. Among other SFR tracers, in analogy with the proposed  [OI]63$\mu$m plus [OIII]88$\mu$m tracer analyzed in \citet{mordini2021}, we suggest the use of the [OIII]88$\mu$m plus [OI]145$\mu$m for high redshift sources. The [OI]145$\mu$m can be detected with ALMA starting from redshift z$\sim$1.5 with observations of $\sim$5\,hr for SFG and LMG-like sources. While in local SFG and LMG there are not enough detections of the [OI]145$\mu$m line to appropriately test it as SFR tracer, this line, although fainter, should be even better than the [OI]63$\mu$m, which can be strongly self-absorbed under certain conditions \citep[see, e.g.,][]{liseau2006} and should allow to measure the SFR in combination with the [OIII]88$\mu$m line.

\item The PDR contribution to the far-IR fine-structure lines in ALMA observations of high-redshift galaxies can be measured with the [CII]/[NII] line ratios, which can be then used to disentangle the PDR and the pure photoionization contributions \citep[see, e.g., ][]{spinoglio2015}.
 
\end{itemize}

{
Finally, as a concluding remark we emphasize the importance of using multi-feature analysis to measure both BH accretion and SFR, since individual tracers can be strongly dependent on the local ISM conditions and vary from source to source. Thus, the combination of more than one tracer provides a more robust determination of these physical quantities.}

\begin{acknowledgement}
We thank the anonymous referee, who helped to improve the paper. LS and JAFO acknowledge financial support by the Agenzia Spaziale Italiana (ASI) under the research contract 2018-31-HH.0. JAFO acknowledges the financial support from the Spanish Ministry of Science and Innovation and the European Union -- NextGenerationEU through the Recovery and Resilience Facility project ICTS-MRR-2021-03-CEFCA.
\end{acknowledgement}


\bibliography{bibliografia}

\end{document}